\documentclass[
aps,
prb,
twocolumn,
superscriptaddress,
longbibliography,
amsmath,
amssymb,
floatfix,
10pt
]{revtex4-2}

\usepackage{graphicx}
\usepackage{color}
\usepackage{xcolor}
\usepackage{comment}

\providecommand{\U}[1]{\protect\rule{.1in}{.1in}}

\begin{document}

\title{Magnetoelectric response to a Néel domain wall in spin–orbit-coupled superconducting films: spontaneous currents, vortex pinning, and vortex–antivortex nucleation}

\author{P. J. F. Cavalcanti}
\email{paulo-jose.fonseca-cavalcanti@u-bordeaux.fr}
\affiliation{University Bordeaux, LOMA UMR-CNRS 5798, F-33405 Talence Cedex, France}
\author{M. D. Croitoru}
\email{mihail.croitoru@ufpe.br}
\affiliation{University Bordeaux, LOMA UMR-CNRS 5798, F-33405 Talence Cedex, France}
\affiliation{Departamento de F\'{\i}sica, Centro de Ci\^encias Exatas e da Natureza, Universidade Federal de Pernambuco, Recife, PE 50740-560, Brazil}
\author{A. I. Buzdin}
\email{alexandre.bouzdine@u-bordeaux.fr}
\affiliation{University Bordeaux, LOMA UMR-CNRS 5798, F-33405 Talence Cedex, France}
\affiliation{Institute for Computer Science and Mathematical Modeling, Sechenov First Moscow State Medical University, Moscow, Russia}

\begin{abstract}
We study the superconducting response induced by a N\'{e}el domain wall in a
thin superconducting film proximity-coupled to a ferromagnetic layer in the
presence of interfacial Rashba spin--orbit coupling. The combined action of
the magnetic texture and broken inversion symmetry generates a
magnetoelectric coupling that induces spontaneous equilibrium currents. To
establish the basic response to a continuously rotating magnetic texture, we
first consider a periodic harmonic cycloid as an analytically tractable
reference problem. Near $T_c$, this reference problem illustrates how the
magnetoelectric coupling modifies the superconducting order parameter, while
below $T_c$ it provides a simple setting for describing the phase response,
Pearl screening, and the interaction with a vortex. We then turn to the main
problem of this work, a single N\'{e}el domain wall, and derive analytical
expressions for its equilibrium phase profile, spontaneous current, magnetic
field, and vortex--wall interaction energy. We show that the N\'{e}el wall
acts as an extended magnetoelectric source and produces a vorticity-selective
pinning landscape controlled by the competition between the wall width and
the effective screening length. We further demonstrate that sufficiently
strong magnetoelectric coupling can render a bound vortex--antivortex pair
energetically favorable relative to the vortex-free state, even in the
absence of an applied magnetic field. Our results provide an analytical
framework for understanding magnetoelectric vortex pinning and
vortex--antivortex nucleation by N\'{e}el domain walls in spin--orbit-coupled
superconducting films.
\end{abstract}

\maketitle

\section{Introduction}

Abrikosov vortices are quantized flux lines in type-II superconductors.
Controlling their pinning, nucleation, motion, and annihilation is central
to superconducting applications and to fluxonics \cite{Vlasko2017,Ge2017}.
Although vortex motion is often associated with dissipation, vortices can also serve as controllable information carriers, motivating the search for mechanisms that allow deterministic control of their position, stability, and dynamics.

A particularly attractive route toward vortex control is provided by
superconductor--ferromagnet hybrid structures \cite%
{Lyuksyutov2005,Aladyshkin2009}, where magnetic textures generate spatially structured magnetic and exchange fields and thereby create effective pinning landscapes for vortices. Early theoretical and experimental work demonstrated that magnetic domain walls can interact with superconductors, modify local electrodynamics, and manipulate individual vortices \cite{Helseth2002,Goa2003}. Related theoretical work has shown that stray-field patterns generated by the distortion of Meissner-current streamlines around nonsuperconducting defects can be used to infer the position and shape of such defects in type-II superconductors \cite{Mironov2016, Plastovets2023}. Subsequent London-theory studies established that Bloch walls may act as efficient magnetic
micromanipulators, producing vortex attraction and guiding effects \cite%
{Vestgarden2007}. More recently, patterned ferromagnetic structures have
been shown to create tunable channels and barriers for vortex motion,
opening routes toward Abrikosov-vortex microcircuits and memory elements 
\cite{Vlasko2017}.

An important aspect of vortex--domain-wall coupling is the possibility of
mutual dynamical control between magnetic textures and superconducting
vortices. Since vortices may become bound to the pinning potential generated by a domain wall, wall motion can induce controlled vortex transport \cite{Vestgarden2007,Diaz2024}. In contrast, vortex supercurrents and magnetic fields may react to the magnetic texture and influence the domain-wall dynamics \cite{Belkin2008,Wilcox2024}. Such reciprocal coupling opens perspectives for hybrid superconducting--magnetic devices with reconfigurable vortex guiding and tunable magnetic textures \cite{Iavarone2014,Fang1990,Kabanov2010,Yang2011,Ishida2021}. In systems with several domain walls, overlapping magnetoelectric pinning landscapes may allow engineered vortex channels and effective potential networks relevant for fluxonics applications and low-dissipation device concepts \cite{Skog2024}. Engineered magnetoelectric landscapes may also generate vortex ratchet effects. In asymmetric or dynamically modulated pinning potentials, alternating drives can produce directional vortex motion and superconducting rectification \cite{Lee1999,Villegas2003,Wordenweber2004,deSouzaSilva2006}. Since domain walls provide spatially extended and potentially reconfigurable vortex potentials, superconductor--ferromagnet hybrids with spin--orbit coupling may offer a natural platform for tunable vortex ratchets and programmable vortex-transport channels \cite{Vlasko2017,Skog2024}.

The control of vortex matter is not limited to the manipulation of
pre-existing vortices, but also includes their controlled nucleation. In
superconducting condensates, vortex--antivortex pairs may emerge under
confinement, inhomogeneous magnetic fields, local heating, or supercurrent
crowding. In particular, Meissner currents flowing around pinning centers
were shown to generate bound vortex dipoles, which can be viewed as
precursors of fully developed vortex--antivortex pairs \cite{Ge2015}. 
A more direct realization of this mechanism was achieved by locally 
suppressing superconductivity with an STM tip, where supercurrent 
crowding around the resulting hot spot leads to controllable 
vortex--antivortex pair generation  \cite{Ge2017}. More recently, 
the on-demand optical generation of individual permanent vortices 
at arbitrary positions inside a superconductor has also been demonstrated, 
based on a rapid quench following the absorption of a tightly 
focused laser pulse that locally drives the system above its critical 
temperature \cite{Rochet2022}. Together, these results show that spatially 
engineered nonequilibrium or current-carrying superconducting states 
can provide a route toward the controlled generation and manipulation of 
superconducting currents and topological excitations
\cite{Croitoru2022Quench,Croitoru2022AQT,Croitoru2023}.

A conceptually different mechanism for controlling superconducting phase and
current distributions arises from spin--orbit coupling (SOC) in systems
without inversion symmetry \cite{Mineev2012,Smidman2017}. In such systems, the superconducting condensate may exhibit magnetoelectric effects: a supercurrent can induce spin polarization, while spin polarization or exchange fields in combination with Rashba SOC can modify the superconducting phase and current response \cite%
{Edelstein1995,Yip2002,Mironov2017}. Within the Ginzburg--Landau framework \cite{Kopnin2001}, this physics is described by Lifshitz invariants, namely, terms linear in the gauge-invariant superconducting momentum and coupled to magnetic or exchange fields \cite{Mineev2007,Mironov2017}.  Experimental signatures of such
Lifshitz-invariant physics have also been observed in synthetic Rashba
superconductors, where an in-plane magnetic field produces an anisotropic
deformation of the vortex response and a corresponding enhancement of vortex
pinning \cite{Fuchs2022}. In homogeneous noncentrosymmetric systems, such terms favor helical superconducting states with finite Cooper-pair momentum \cite{Edelstein1989,Samokhin2004,Kaur2005,Agterberg2007,Dimitrova2008}, which are similar to the Fulde--Ferrell--Larkin--Ovchinnikov (FFLO) states \cite{Larkin1964,Fulde1964,Croitoru2012b,Croitoru2016,Croitoru2017}. However, a finite Cooper-pair momentum does not by itself imply an uncompensated charge current: in uniform systems the anomalous spin--orbit contribution to the current may cancel the conventional orbital contribution \cite{Mineev1994}. In contrast, at superconductor--ferromagnet interfaces with local Rashba SOC and an exchange field, this cancellation is avoided, giving rise to spontaneous interfacial currents and the associated screening response \cite{Mironov2017}.

At such interfaces, broken structural inversion symmetry naturally generates Rashba SOC, so magnetoelectric effects become particularly important \cite{Bychkov1984,Gorkov2001,Pershoguba2015,Hals2016,Mironov2017,Plastovets2024,Melnikov2022}.
Interfacial SOC can induce spontaneous vortices in
superconducting films in contact with ferromagnets, even in zero applied
magnetic field \cite{Olde2019}. Similarly, magnetic skyrmions may generate
spontaneous currents and nucleate superconducting vortices through the same magnetoelectric mechanism \cite{Baumard2019}. More recently, SOC has been shown to strongly modify vortex--domain-wall interactions, allowing domain walls to push, drag, or guide vortices depending on their helicity \cite{Diaz2024}. These results establish magnetic textures as active sources of superconducting currents and vortex potentials.

A periodically rotating planar texture provides a useful reference geometry
for analyzing the magnetoelectric response generated by a continuously
varying magnetization.\cite{Lautenschlager1993,Arkenbout2006,Seki2008,Adler2022,
Dashwood2020,Faure2023} Cycloidal magnetic order, in which the magnetic
moments rotate continuously along a propagation direction lying in the
spin-rotation plane, occurs in several magnetic materials, including
TbMnO$_3$, DyMnO$_3$, Ni$_3$V$_2$O$_8$, and BiFeO$_3$
\cite{Kenzelmann2005,Finger2014,Kenzelmann2006,Cabrera2009,
Bordacs2018,Meisenheimer2024}. In the following, however, the periodic
harmonic cycloid is used primarily as an analytically transparent reference
problem for establishing the superconducting response to a rotating
magnetoelectric source before considering the isolated N\'{e}el domain wall.



A N\'{e}el domain wall provides a particularly interesting realization of the magnetoelectric mechanism. Unlike sharp magnetic interfaces or
localized magnetic textures, the magnetization rotates continuously over the wall region, generating a distributed source of superconducting phase
gradients and spontaneous currents. The resulting electrodynamic response in a thin superconducting film is governed by Pearl screening \cite{Pearl1964} and is therefore controlled by the interplay between the wall geometry and the effective screening length. Thus, the spatial profile of the magnetic texture is directly encoded in the resulting vortex pinning landscape.

In this work, our main objective is to determine the superconducting
electrodynamic response to a single N\'{e}el domain wall whose magnetization
rotates over a region of width $2w$. We treat the wall as an extended
interfacial magnetoelectric source and determine the resulting phase,
spontaneous current, magnetic field, vortex pinning potential, and conditions
for vortex--antivortex pair formation. Before addressing the isolated wall,
we consider a periodic harmonic cycloid as a reference problem. This
periodic geometry allows the basic consequences of a continuously rotating
magnetoelectric source to be established in a particularly transparent
form. Near $T_c$, it illustrates the modulation of the superconducting
order parameter at the transition, while below $T_c$ it provides a simple
example of the equilibrium phase response, Pearl-screened current and
magnetic field, and vortex interaction. We then apply the same
magnetoelectric and London/Pearl framework to the single N\'{e}el wall,
which is the central geometry considered in this work. The wall produces an
extended and vorticity-selective pinning landscape and modifies the energy
of vortex--antivortex configurations. Within the logarithmic approximation
for the vortex--antivortex interaction used in the pair-formation criterion,
sufficiently strong magnetoelectric coupling can make a bound
vortex--antivortex pair energetically favorable even in the absence of an
applied magnetic field.
Our results provide an analytical framework for understanding how a
N\'{e}el domain wall generates magnetoelectric currents and magnetic fields
and how the resulting electrodynamic response controls vortex pinning and
vortex--antivortex pair formation.

The paper is organized as follows. In Sec.~\ref{sec:geometry}, we introduce
the geometry of the system and the magnetoelectric source generated by the
magnetic texture. In Sec.~\ref{sec:harmonic}, we use a periodic harmonic
cycloid as a reference problem to establish the superconducting response to
a continuously rotating magnetoelectric source near and below $T_c$.
We then turn to the main problem of the paper, the isolated N\'{e}el wall.
In Sec.~\ref{sec:model}, we derive its equilibrium phase profile and
spontaneous current, while Sec.~\ref{sec:london} gives the Pearl-screened
current, magnetic field, and vortex--wall interaction energy. In
Sec.~\ref{sec:pairs}, we use this interaction to formulate the energetic
criterion for stabilizing a bound vortex--antivortex pair. Finally,
Sec.~\ref{sec:conclusions} summarizes the main results. Additional
derivations and technical details are provided in the Supplemental
Material~\cite{SupplementalMaterial}.

\section{System geometry and magnetoelectric source}
\label{sec:geometry}

We consider a thin superconducting film lying in the plane $z=0$ and coupled to a ferromagnetic layer of thickness $d_f$ on top. The magnetization depends only on $x$ and forms a single N\'{e}el domain wall centered at $x=0$, with the rotation occurring over the interval $\left\vert x\right\vert \leqslant w $. Far from the wall the magnetization is uniform, while inside the wall it rotates continuously in the plane. The interface normal is taken as $\hat{\mathbf{n}}=\hat{\mathbf{z}}$. 

In the standard one-dimensional micromagnetic description, a smooth N\'{e}el-wall profile may be written as $\mathbf{M}(x)=m(\cos\theta,\sin\theta,0)$, with $\sin\theta=\tanh\!\left(\frac{x}{2w}\right)$ and $\cos\theta=\operatorname{sech}\!\left(\frac{x}{2w}\right)$, where $m$ denotes the magnitude of the in-plane magnetization \cite{Aharoni1966,Cervera2004,Diaz2024}. For analytical tractability, we replace the smooth micromagnetic profile by a single-wall texture in which the magnetization rotates uniformly over the interval $\left\vert x\right\vert \leqslant w $ and matches continuously onto the two homogeneous domains:
\begin{equation}
\mathbf{M}(x)=m
\begin{cases}
(0,-1,0), & x<-w,\\[4pt]
\left[\cos\!\left(\frac{\pi x}{2w}\right),\,\sin\!\left(\frac{\pi x}{2w}\right),\,0\right], & |x|\le w,\\[4pt]
(0,+1,0), & x>w.
\end{cases}
\label{eq:M-piecewise}
\end{equation}

Throughout this work, we use Gaussian (cgs) units. We focus on the
spin--orbit-driven magnetoelectric response. The magnetic field generated by the magnetoelectrically induced superconducting current is retained self-consistently through Pearl screening.

We describe the film by the thickness-averaged free-energy density
$f_{\mathrm{sc}}$, so that the total free energy is

\begin{equation}
F = d_s\int d^2r\, f_{\mathrm{sc}}
+
\int d^3r\, \frac{\mathbf{h}^2}{8\pi},
\end{equation}
where $d_s$ is the superconducting-film thickness and
$\mathbf{h}=\nabla\times\mathbf{A}$ is the magnetic induction in the
surrounding three-dimensional space. 

We start from the Ginzburg--Landau functional \cite{Mironov2017}
\begin{equation}
\begin{aligned}
f_{\mathrm{sc}}
={}&
a|\psi|^2
+
\frac{b}{2}|\psi|^4
+
\gamma|\mathbf{D}\psi|^2
\\
&+
\varepsilon
\left\{
\psi^*\big[(\mathbf{n}\times\mathbf{M})\cdot \mathbf{D}\big]\psi
+\mathrm{c.c.}
\right\}.
\end{aligned}
\label{eq:gl-functional}
\end{equation}
Here $a=a_0(T-T_c^0)$ and $b$ are the standard Ginzburg--Landau coefficients, with $T_c^0$ the bare critical temperature, $\gamma$ is the stiffness parameter, $\psi$ is the superconducting order parameter, and $\mathbf{A}$ is the vector potential. The covariant derivative is $\mathbf{D}=-i\nabla-\frac{2e}{\hbar c}\mathbf{A}$, with $-e$ the electron charge and $c$ the speed of light. The unit vector $\mathbf{n}$ is normal to the film. The last term is a Lifshitz invariant describing the interfacial magnetoelectric coupling generated by the combined action of Rashba SOC and the in-plane magnetization.

In a homogeneous region, a Lifshitz invariant of this form favors a
helical superconducting state with a uniform phase gradient,
$\psi(\mathbf{r})=\psi_0e^{i\mathbf{q}\cdot\mathbf{r}}$, where
$\mathbf{q}$ is the helical wave vector
\cite{Kaur2005,Mineev2007,Mironov2017,Paulo2026}.
Neglecting the direct orbital field and setting $\mathbf{A}=0$,
substitution of this ansatz into Eq.~(\ref{eq:gl-functional}) gives
\begin{equation}
f(\mathbf{q})=
a|\psi_0|^2
+
\frac{b}{2}|\psi_0|^4
+
\gamma|\psi_0|^2 q^2
+
2\varepsilon|\psi_0|^2(\mathbf{n}\times\mathbf{M})\cdot\mathbf{q}.
\label{eq:fq_complete}
\end{equation}
This expression makes explicit that the magnetoelectric coupling favors a helical modulation of the superconducting phase. Minimization with respect to $\mathbf{q}$ yields
\begin{equation}
\mathbf{q}=-\frac{\varepsilon}{\gamma}(\mathbf{n}\times\mathbf{M}).
\label{eq:q0}
\end{equation}
Thus, the magnetoelectric coupling selects a helical superconducting state whose preferred phase gradient is fixed by the magnetization orientation. In a homogeneous region and in the absence of an orbital field, the conventional phase-gradient contribution to the current is then exactly compensated by the anomalous magnetoelectric contribution, so that the equilibrium charge current vanishes.

It is therefore convenient to introduce the corresponding magnetoelectric source field
\begin{equation}
\boldsymbol{\alpha}(x)
=
\alpha_0\,\frac{\mathbf{n}\times\mathbf{M}(x)}{m},
\qquad
\alpha_0=\frac{\varepsilon m}{\gamma}.
\label{eq:alpha-def}
\end{equation}
For a homogeneous magnetization, the locally preferred condensate momentum is simply $\mathbf{q}=-\boldsymbol{\alpha}$.
For a spatially varying magnetic texture, however, $\boldsymbol{\alpha}(x)$ varies in space and acts as a distributed magnetoelectric source. The superconducting phase can compensate its longitudinal component through an equilibrium phase gradient, whereas the uncompensated transverse component generates a spontaneous current. This distinction is central to the electrodynamic response derived below.

For the N\'{e}el-wall profile, with $\mathbf{n}=\hat{\mathbf{z}}$, this becomes
\begin{equation}
\boldsymbol{\alpha}(x)=\alpha_0
\begin{cases}
(+1,0,0), & x<-w,\\[4pt]
\left(-\sin\!\left(\frac{\pi x}{2w}\right),\,\cos\!\left(\frac{\pi x}{2w}\right),\,0\right), & |x|\le w,\\[4pt]
(-1,0,0), & x>w.
\end{cases}
\label{eq:alpha-piecewise}
\end{equation}
The longitudinal component $\alpha_x(x)$ determines the equilibrium phase gradient, while the transverse component $\alpha_y(x)$ acts as the source of the spontaneous current flowing parallel to the wall.

\begin{figure}[tb]
\centering \includegraphics[width=%
\columnwidth]{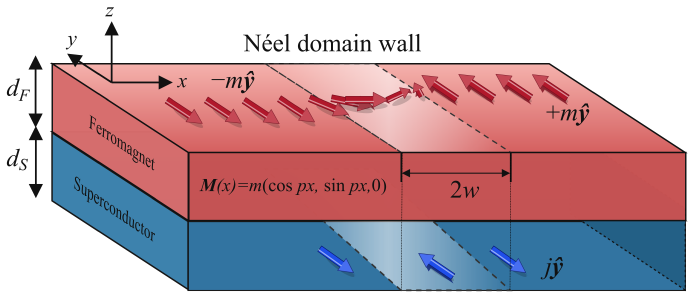}
\caption{Thin superconducting film at $z=0$ with a thin ferromagnet on top;
magnetization depends only on $x$ and forms a N\'{e}el domain wall of width $%
2w$ centered at $x=0$.}
\label{fig:01}
\end{figure}

Figure~\ref{fig:01} summarizes the geometry. The N\'{e}el wall extends along the $y$ direction, while both the magnetic texture and the magnetoelectric source vary only along $x$. For the single-wall
equilibrium problem, translational invariance along $y$, together with
the current-free asymptotic conditions in the homogeneous domains,
allows the London phase to be chosen as $\varphi=\varphi(x)$. This one-dimensional structure determines the symmetry of the spontaneous current, magnetic field, and vortex pinning potential derived below.

A microscopic estimate for an interfacial Rashba layer gives
\begin{equation}
|\alpha_0|
\sim
\left(\frac{v_R}{v_F}\right)
\left(\frac{a_{\mathrm{lat}}}{d_s}\right)
\left(\frac{h_{\mathrm{ex}}}{k_{\mathrm{B}}T_c}\right)
\frac{1}{\xi},
\label{eq:alpha0-microscopic}
\end{equation}
where $v_R$ and $v_F$ are the Rashba and Fermi velocities,
$a_{\mathrm{lat}}$ is an atomic interfacial length, and $h_{\mathrm{ex}}$ is the
exchange-energy scale \cite{Olde2019}. Representative ratios
$v_R/v_F\sim10^{-1}$, $a_{\mathrm{lat}}/d_s\sim10^{-1}$, and
$h_{\mathrm{ex}}/(k_{\mathrm{B}}T_c)\sim1$ give $|\alpha_0|\xi\sim10^{-2}$.
Thus, the local helical wave vector remains within the gradient-expansion
regime, $|\mathbf{q}|\xi\sim|\alpha_0|\xi\ll1$.

Note that the above estimate corresponds to the clean-limit treatment, when
the electrons mean free path $\ell \gtrsim \xi, d_{s}$. In the dirty limit, $\ell \ll \xi$, 
impurity scattering renormalizes the magnetoelectric coupling, but does not suppress it completely. 
For a two-dimensional Rashba superconductor, quasiclassical calculations show that 
the coefficient $|\alpha_0|$ decreases by a factor $\ell / \xi$ \cite{Houzet2015}. 
Therefore, the value of $|\alpha_0|$ used below should be regarded as an order-of-magnitude estimate. 
For realistic interfaces, the magnetoelectric coupling may in principle be determined 
from a material-specific relativistic Bogoliubov--de Gennes calculation, such as that of Ref. \cite{Csire2018}.

For the thin-film electrodynamics used below, we assume $d_s\ll\lambda_L$ and $d_s\ll w$, so that the film can be represented by a sheet current at the wall scale. We introduce the effective screening length
$\lambda_{\mathrm{eff}}=\lambda_L^2/d_s$, where $\lambda_L^2=\hbar^2c^2/(32\pi e^2\gamma\psi_0^2)$ and $\lambda_L$ is the London penetration depth. We also use $\phi_0=\pi\hbar c/e$ for the flux quantum and denote by $\xi$ the Ginzburg--Landau coherence length. The representative thin-film values $\xi/\lambda_{\mathrm{eff}}\sim10^{-3}-10^{-2}$.

In our approach, we restrict the discussion to domain walls with half-width $w \gg \xi$, 
while broader walls are described by varying $w/\lambda_{\mathrm{eff}}$ explicitly. These parameter 
ranges are consistent with those commonly considered for magnetoelectric effects, 
spontaneous currents, and texture-induced vortices in spin--orbit-coupled 
superconductor--ferromagnet structures \cite{Mironov2017,Olde2019,Baumard2019,Plastovets2024}.

Before solving the electrodynamic problem for the isolated wall, it is useful
to consider a periodic continuation of its rotating part. This reference
problem isolates the response to a continuously rotating magnetoelectric
source and permits several results to be obtained in a particularly simple
analytical form. The single-wall problem is then treated in
Secs.~\ref{sec:model}--\ref{sec:pairs}.

\section{Periodic cycloid as a reference problem}

\label{sec:harmonic}

We first consider a periodic harmonic cycloid in order to establish, in a
simple reference geometry, the superconducting response to a continuously
rotating magnetoelectric source. The rotating part of this periodic texture
is identical to that of the single N\'{e}el wall introduced in
Sec.~\ref{sec:geometry}. The periodicity makes the near-$T_c$ problem
amenable to a Fourier-mode analysis and, below $T_c$, reduces the
London/Pearl electrodynamics to a single spatial harmonic. These results
provide a useful benchmark before we return in Sec.~\ref{sec:model} to the
isolated N\'{e}el wall, which is the main geometry of interest.

\subsection{Linearized Ginzburg--Landau analysis near $T_c$}

Close to the superconducting transition, the order
parameter is determined by the linearized Ginzburg--Landau equation.
At the transition, this equation first admits a nontrivial solution, which determines the order-parameter profile at the onset of superconductivity. We consider a periodic harmonic cycloid. Its magnetization rotates in the film plane, while the magnetic modulation
propagates along the $x$ direction, so that the propagation direction lies
in the spin-rotation plane. In terms of the magnetoelectric source introduced in Sec.~\ref{sec:geometry}, we take
\begin{equation}
\boldsymbol{\alpha}_{\mathrm{cyc}}(x)
=
\alpha_0
\left[
-\sin\!\left(\frac{\pi x}{2w}\right),
\cos\!\left(\frac{\pi x}{2w}\right),0
\right].
\label{eq:alpha-inner-harmonic}
\end{equation}
For $|x|\le w$, this source coincides with the rotating branch of the single-wall source in Eq.~(\ref{eq:alpha-piecewise}). 

Varying Eq.~(\ref{eq:gl-functional}) with respect to $\psi^*$ gives
\begin{equation}
 a\psi+b|\psi|^2\psi+\gamma\mathbf{D}^2\psi
+\gamma\left[
2\boldsymbol{\alpha}\cdot\mathbf{D}\psi
-i(\nabla\cdot\boldsymbol{\alpha})\psi
\right]=0.
\label{eq:gl-alpha-full}
\end{equation}
The divergence term is absent for a uniform helical source, but it is essential for a spatially rotating magnetoelectric texture. In the linearized problem, the magnetic field generated by the incipient condensate is of order $|\psi|^2$ and therefore does not enter the eigenvalue equation. We therefore set $\mathbf{A}=0$ and determine the order-parameter profile induced by the periodic magnetoelectric source at the onset of superconductivity. The direct orbital coupling to the ferromagnetic stray field is compared with the Lifshitz source in
Sec.~\ref{sec:harmonic-london}, where the SOC-dominated regime is specified.

Since the cycloidal source is translationally invariant
along $y$, the eigenmodes can be written as
$\psi(x,y)=\psi_{q_y}(x)e^{iq_y y}$, with
$\psi_{q_y}(x+4w)=\psi_{q_y}(x)$. At the superconducting onset, the
physical state is the mode with the highest transition temperature. In
the weak-magnetoelectric regime
$2|\alpha_0|w/\pi\ll1$ considered below, perturbative minimization of the
lowest eigenvalue with respect to $q_y$ selects $q_y=0$. We therefore use
$\psi(x,y)=\psi(x)$, with $\psi(x+4w)=\psi(x)$.
The linearized Ginzburg--Landau equation then reduces to
\begin{multline}
0=a\psi
-\gamma\frac{d^2\psi}{dx^2}
+i\gamma\alpha_0
\Bigg[
2\sin\!\left(\frac{\pi x}{2w}\right)\frac{d\psi}{dx}
\\
+\frac{\pi}{2w}\cos\!\left(\frac{\pi x}{2w}\right)\psi
\Bigg].
\label{eq:gl-linear-nearTc}
\end{multline}
Because the coefficients are periodic with period $4w$, we expand
\begin{equation}
\psi(x)=\sum_{n\in\mathbb{Z}}\psi_n
\exp\!\left(i n\frac{\pi x}{2w}\right).
\label{eq:fourier-nearTc}
\end{equation}
Projecting Eq.~(\ref{eq:gl-linear-nearTc}) onto
$\exp[i\ell\pi x/(2w)]$ gives
\begin{equation}
\begin{aligned}
0={}&
\left[
 a+\gamma\left(\frac{\pi}{2w}\right)^2\ell^2
\right]\psi_\ell
\\
&+
 i\gamma\alpha_0\left(\frac{\pi}{2w}\right)
\left[
\left(\ell-\frac12\right)\psi_{\ell-1}
-
\left(\ell+\frac12\right)\psi_{\ell+1}
\right].
\end{aligned}
\label{eq:recurrence-nearTc}
\end{equation}
Thus, the rotating magnetoelectric source couples only neighboring Fourier
harmonics.

The periodic source couples the uniform mode $\psi_0$ directly only to the
first Fourier pair $\psi_{\pm1}$; the $\psi_{\pm2}$ amplitudes arise only
through a second successive coupling. The resulting leading $0,\pm1$
subsystem gives
\begin{equation}
\psi_{\pm1}
=
-i\,
\frac{\gamma\alpha_0(\pi/2w)}
{2\left[a+\gamma(\pi/2w)^2\right]}
\psi_0,
\end{equation}
and the corresponding solvability condition within the leading
$0,\pm1$ truncation is
\begin{equation}
2a\left[a+\gamma\left(\frac{\pi}{2w}\right)^2\right]
-
\gamma^2\alpha_0^2\left(\frac{\pi}{2w}\right)^2=0.
\label{eq:critical-nearTc}
\end{equation}
Within the $0,\pm1$ truncation, the root of Eq.~(\ref{eq:critical-nearTc}) that is continuously connected to $a=0$ in the limit $\alpha_0\to0$ determines the critical value $a_c$ to leading order and hence gives the leading shift of the superconducting transition. Higher Fourier harmonics do not modify this leading contribution, but the first relative correction in
$(2\alpha_0w/\pi)^2$ requires inclusion of the $\ell=\pm2$ Fourier components.

In the weak-magnetoelectric regime $2|\alpha_0|w/\pi\ll1$, the
recurrence relation generates a perturbative hierarchy of Fourier amplitudes. The first pair obeys $\psi_{\pm1}/\psi_0=
\mathcal{O}(2\alpha_0w/\pi)$, while the next Fourier pair appears only at second order. To leading nonvanishing order, the $\ell=\pm2$ components obey
\begin{equation}
\psi_{\pm2}
=
-i\,
\frac{3\gamma\alpha_0(\pi/2w)}
{2\left[a_c+4\gamma(\pi/2w)^2\right]}
\psi_{\pm1}.
\label{eq:psi2-nearTc}
\end{equation}
Consequently,
\begin{equation}
\left.
\left|
\frac{\psi_{\pm2}}{\psi_{\pm1}}
\right|
\right|_{a=a_c}
=
\frac{3|\alpha_0|w}{4\pi}
+
\mathcal{O}\!\left[
\left(\frac{2|\alpha_0|w}{\pi}\right)^3
\right].
\label{eq:first-harmonic-validity-main}
\end{equation}
It follows that $\psi_{\pm2}/\psi_0=
\mathcal{O}[(2\alpha_0w/\pi)^2]$. The uniform component and the first Fourier pair therefore constitute the
leading-order eigenfunction at the transition, while
higher Fourier pairs enter successively at higher orders in the
magnetoelectric coupling.
Substitution of Eq.~(\ref{eq:psi2-nearTc}) into the
$\ell=\pm1$ recurrence relations, followed by the $\ell=0$ solvability
condition, gives
\begin{equation}
a_c=
\frac{\gamma\alpha_0^2}{2}
\left[
1+\frac{1}{16}\left(\frac{2\alpha_0w}{\pi}\right)^2
+\mathcal{O}\!\left[\left(\frac{2\alpha_0w}{\pi}\right)^4\right]
\right].
\label{eq:ac-nearTc}
\end{equation}
With $a=a_0(T-T_c^0)$, the corresponding critical
temperature is

\begin{equation}
\begin{aligned}
T_{c,\mathrm{cyc}}
={}&
T_c^0+
\frac{\gamma\alpha_0^2}{2a_0}
\\
&\times
\left[
1+\frac{1}{16}\left(\frac{2\alpha_0w}{\pi}\right)^2
+\mathcal{O}\!\left[
\left(\frac{2\alpha_0w}{\pi}\right)^4
\right]
\right].
\end{aligned}
\label{eq:TcN-nearTc}
\end{equation}
Thus, the leading magnetoelectric enhancement of the transition temperature is $\gamma\alpha_0^2/(2a_0)$, while the correction proportional to $(2\alpha_0w/\pi)^2$ in Eq.~(\ref{eq:TcN-nearTc}) arises from coupling to the second Fourier pair.

At $a=a_c$, the rotating magnetoelectric texture mixes the uniform condensate
component with the first spatial Fourier pair. In the weak-magnetoelectric
regime, the resulting leading correction is a phase modulation. Up to an arbitrary global phase, the critical order-parameter profile is

\begin{equation}
\psi_{\mathrm{cyc}}(x)=
\psi_0
\left[
1
-i\frac{2\alpha_0w}{\pi}
\cos\!\left(\frac{\pi x}{2w}\right)
+
\mathcal{O}\!\left[
\left(\frac{2\alpha_0w}{\pi}\right)^2
\right]
\right].
\label{eq:eigenmode-nearTc}
\end{equation}
The linearized equation fixes the relative Fourier weights, whereas the
magnitude of $\psi_0$ is determined only when the nonlinear term is retained
below the transition. The correction of relative order
$(2\alpha_0w/\pi)^2$ contains the second Fourier pair and is therefore beyond
the three-mode approximation. To the same order of control, the phase of
$\psi_{\mathrm{cyc}}(x)$ is
\begin{equation}
\varphi_{\mathrm{cyc}}(x)=
-\frac{2\alpha_0w}{\pi}
\cos\!\left(\frac{\pi x}{2w}\right)
+
\mathcal{O}\!\left[\left(\frac{2\alpha_0w}{\pi}\right)^3\right].
\label{eq:phase-nearTc}
\end{equation}

\subsection{London electrodynamics and vortex interaction below $T_c$}
\label{sec:harmonic-london}

Below $T_c$, we treat the periodic cycloid within the
constant-modulus London approximation,
$\psi_{\mathrm{cyc}}=\psi_0e^{i\varphi_{\mathrm{cyc}}}$. The longitudinal
equilibrium condition
$\partial_x\varphi_{\mathrm{cyc}}+\alpha_{\mathrm{cyc},x}=0$
implies that, within this approximation, the phase is given exactly by the
leading cosine term in Eq.~(\ref{eq:phase-nearTc}). The transverse component
remains
$\alpha_{\mathrm{cyc},y}=\alpha_0\cos(\pi x/2w)$.

The current density in the constant-modulus regime is
\begin{equation}
\mathbf{j}=
\frac{2e\gamma}{\hbar}
\left[
 i\psi\nabla\psi^*-i\psi^*\nabla\psi
+2\left(\boldsymbol{\alpha}-\frac{2e}{\hbar c}\mathbf{A}\right)|\psi|^2
\right].
\label{eq:current-general-harmonic}
\end{equation}
Using $2e/(\hbar c)=2\pi/\phi_0$ and the definition of $\lambda_L$,
Eq.~(\ref{eq:current-general-harmonic}) becomes
\begin{equation}
\mathbf{j}
=
-\frac{c}{4\pi\lambda_L^2}
\left[
\mathbf{A}
-
\frac{\phi_0}{2\pi}
\left(\nabla\varphi+\boldsymbol{\alpha}\right)
\right].
\label{eq:london-current-source}
\end{equation}
Thus, the magnetoelectric field enters the London combination through
$(\phi_0/2\pi)\boldsymbol{\alpha}$. Upon integration across the film
thickness, the sheet-current scale is
\begin{equation}
j_0
\equiv
d_s\frac{4e}{\hbar}\gamma\psi_0^2\alpha_0.
\label{eq:j0-definition}
\end{equation}
Using the longitudinal equilibrium condition
$\partial_x\varphi_{\mathrm{cyc}}+\alpha_{\mathrm{cyc},x}=0$,
the unscreened sheet current is
\begin{equation}
 j_{\mathrm{cyc},x}^{(0)}=0,
\qquad
 j_{\mathrm{cyc},y}^{(0)}(x)
 =
 j_0\cos\!\left(\frac{\pi x}{2w}\right).
\label{eq:leading-current-harmonic}
\end{equation}
Equation~(\ref{eq:leading-current-harmonic}) is exact within the
constant-modulus London description of the prescribed harmonic source.
The near-$T_c$ Fourier expansion and the London calculation address
complementary regimes: the former is perturbatively controlled for
$2|\alpha_0|w/\pi\ll1$, whereas the latter treats the prescribed texture
without a truncation of its spatial profile.

The periodic cycloidal magnetization,
$\mathbf{M}_{\mathrm{cyc}}(x)=m[\cos(\pi x/2w),\sin(\pi x/2w),0]$, carries
the magnetic-charge density
$\rho_m=-\nabla\cdot\mathbf{M}_{\mathrm{cyc}}=(\pi m/2w)\sin(\pi x/2w)$.
In the thin-ferromagnet limit $d_f\ll w$, treating the magnetic layer as an
interfacial sheet, the stray field and the corresponding vector potential
on the superconducting side, $z<0$, are
\begin{equation}
\begin{aligned}
\mathbf{h}^{\mathrm{M}}(x,z)
&=
-\frac{\pi^2m d_f}{w}
\exp\!\left(\frac{\pi z}{2w}\right)
\Bigg[
\cos\!\left(\frac{\pi x}{2w}\right)\hat{\mathbf{x}}
\\[-2pt]
&\hspace{4.0em}
+
\sin\!\left(\frac{\pi x}{2w}\right)\hat{\mathbf{z}}
\Bigg],
\\
A_y^{\mathrm{M}}(x,z)
&=
2\pi m d_f
\cos\!\left(\frac{\pi x}{2w}\right)
\exp\!\left(\frac{\pi z}{2w}\right).
\end{aligned}
\label{eq:magnetostatic-harmonic-source}
\end{equation}

At the superconducting plane, the
magnetostatic vector potential is
$A_y^{\mathrm M}(x,0^-)=2\pi m d_f\cos(\pi x/2w)$. Since both contributions have the same spatial harmonic, the combined
transverse source driving the superconducting response is
\begin{equation}
\begin{aligned}
\mathcal{S}_y^{\mathrm{eff}}(x)
&=
\frac{\phi_0}{2\pi}\alpha_{\mathrm{cyc},y}(x)
-
A_y^{\mathrm M}(x,0^-)
\\
&=
\left[
\frac{\phi_0\alpha_0}{2\pi}
-
2\pi m d_f
\right]
\cos\!\left(\frac{\pi x}{2w}\right).
\end{aligned}
\label{eq:Seff-harmonic}
\end{equation}
The spin--orbit and direct orbital contributions may therefore reinforce or
partially compensate each other, depending on their relative sign, which is
controlled by the sign of $\alpha_0$ and by the orientation of the magnetic
texture relative to the interface normal.

To keep both contributions explicit in the electrodynamic response, we write
Eq.~(\ref{eq:Seff-harmonic}) as
$\mathcal{S}_y^{\mathrm{eff}}(x)
=
\mathcal{S}_0^{\mathrm{eff}}
\cos(\pi x/2w)$, where
$\mathcal{S}_0^{\mathrm{eff}}
=
\phi_0\alpha_0/(2\pi)-2\pi m d_f$.
Since the effective source contains only the harmonic with wave number
$\pi/(2w)$, Pearl screening preserves its in-plane spatial dependence, while
the vector potential decays away from the film as
$\exp[-(\pi/2w)|z|]$. The thin-film Pearl denominator is therefore
$1+\pi\lambda_{\mathrm{eff}}/w$, and the vector potential at the film plane
is
\begin{equation}
A_{\mathrm{cyc},y}(x,0)
=
\frac{\mathcal{S}_0^{\mathrm{eff}}}
{1+\pi\lambda_{\mathrm{eff}}/w}
\cos\!\left(\frac{\pi x}{2w}\right).
\label{eq:Ay-harmonic}
\end{equation}
Here and below, $A_{\mathrm{cyc},y}$ denotes the vector potential generated
by the superconducting screening response; the direct magnetostatic
contribution $A_y^{\mathrm M}$ has already been included in
$\mathcal{S}_y^{\mathrm{eff}}$.

Thus, for a single harmonic, Pearl screening preserves the spatial profile
of the vector potential and reduces its amplitude by the factor
$(1+\pi\lambda_{\mathrm{eff}}/w)^{-1}$. At the film plane, the perpendicular
magnetic field is obtained directly from
$h_{\mathrm{cyc},z}(x)=\partial_xA_{\mathrm{cyc},y}(x,0)$. Using
Eq.~(\ref{eq:Ay-harmonic}) gives
\begin{equation}
h_{\mathrm{cyc},z}(x)
=
-
\frac{\pi\mathcal{S}_0^{\mathrm{eff}}}{2w}
\frac{
\sin\!\left(\frac{\pi x}{2w}\right)}
{1+\pi\lambda_{\mathrm{eff}}/w}.
\label{eq:hz-harmonic}
\end{equation}

The screened sheet current follows from the transverse London relation
\begin{equation}
j_{\mathrm{cyc},y}(x)
=
\frac{c}{4\pi\lambda_{\mathrm{eff}}}
\left[
\mathcal{S}_y^{\mathrm{eff}}(x)
-
A_{\mathrm{cyc},y}(x,0)
\right].
\label{eq:jcyc-london-screened}
\end{equation}
Substituting Eq.~(\ref{eq:Seff-harmonic}) and
Eq.~(\ref{eq:Ay-harmonic}) into
Eq.~(\ref{eq:jcyc-london-screened}) yields
\begin{equation}
j_{\mathrm{cyc},y}(x)
=
\frac{c\mathcal{S}_0^{\mathrm{eff}}}
{4\pi\lambda_{\mathrm{eff}}}
\frac{\pi\lambda_{\mathrm{eff}}/w}
{1+\pi\lambda_{\mathrm{eff}}/w}
\cos\!\left(\frac{\pi x}{2w}\right).
\label{eq:screened-current-harmonic}
\end{equation}
Equations~(\ref{eq:Ay-harmonic})--(\ref{eq:screened-current-harmonic})
show explicitly that the spin--orbit and direct orbital contributions enter
the Pearl-screened response through the same effective source
$\mathcal{S}_0^{\mathrm{eff}}$.

In the SOC-dominated regime considered in the following, the direct orbital
contribution can be neglected when
$2\pi m d_f\ll|\alpha_0|\phi_0/(2\pi)$, so that
\begin{equation}
\mathcal{S}_0^{\mathrm{eff}}
\simeq
\frac{\phi_0\alpha_0}{2\pi}.
\label{eq:Seff-SOC-limit}
\end{equation}
We use this inequality as an order-of-magnitude criterion for the
SOC-dominated regime in the single-wall analysis below. Using
$H_{c2}=\phi_0/(2\pi\xi^2)$, the condition becomes
$4\pi m\ll2(|\alpha_0|\xi)H_{c2}\xi/d_f$. For
$d_f\sim2|\alpha_0|\xi^2$, it is parametrically equivalent to
$4\pi m\ll H_{c2}$; for thinner ferromagnets, the latter is a sufficient
conservative condition.

In this limit, Eq.~(\ref{eq:screened-current-harmonic}) reduces to
\begin{equation}
\frac{j_{\mathrm{cyc},y}(x)}{j_0}
=
\frac{\pi\lambda_{\mathrm{eff}}/w}
{1+\pi\lambda_{\mathrm{eff}}/w}
\cos\!\left(\frac{\pi x}{2w}\right),
\label{eq:screened-current-harmonic-SOC}
\end{equation}
which is the form used below for the vortex interaction.

We now compute the interaction with a Pearl vortex of topological charge $s=\pm1$ located at $\mathbf R_0=(x_0,y_0)$ from the mixed term in the
London/Pearl energy. In the constant-modulus regime, the regular source generated by the cycloid is
\begin{equation}
\mathbf C_{\mathrm{cyc}}(\mathbf r)
=
\frac{\phi_0}{2\pi}
\left[
\nabla\varphi_{\mathrm{cyc}}(\mathbf r)
+
\boldsymbol{\alpha}_{\mathrm{cyc}}(\mathbf r)
\right],
\label{eq:Ccyc-def}
\end{equation}
while the vortex source is
\begin{equation}
\boldsymbol{\Phi}^{(s)}(\mathbf r)
=
\frac{\phi_0}{2\pi}\nabla\varphi_{\mathrm{vor}}^{(s)}(\mathbf r).
\label{eq:Phi-vor-def}
\end{equation}
The corresponding interaction energy is
\begin{equation}
\begin{aligned}
E_{\mathrm{int},\mathrm{cyc}}^{(s)}
={}&
\frac{1}{4\pi}
\int\frac{d^2k}{(2\pi)^2}
\frac{|\mathbf{k}|}
{1+2\lambda_{\mathrm{eff}}|\mathbf{k}|}
\\
&\times
\left[
\mathbf{C}_{\mathrm{cyc},\mathbf{k}}
\cdot
\boldsymbol{\Phi}^{(s)*}_{\mathbf{k}}
+
\mathrm{c.c.}
\right].
\end{aligned}
\label{eq:eint-cyc-general}
\end{equation}

For the harmonic cycloid,
$\nabla\varphi_{\mathrm{cyc}}+\boldsymbol{\alpha}_{\mathrm{cyc}}
=\alpha_0\cos(\pi x/2w)\hat{\mathbf y}$, so that only the transverse
harmonic components contribute. The vortex source in Fourier space is
\begin{equation}
\boldsymbol{\Phi}^{(s)}_{\mathbf k}
=
is\phi_0
e^{-i\mathbf k\cdot\mathbf R_0}
\frac{
k_y\hat{\mathbf x}-k_x\hat{\mathbf y}}
{|\mathbf k|^2}.
\label{eq:Phi-vortex-fourier-harmonic}
\end{equation}
Substitution of these sources into Eq.~(\ref{eq:eint-cyc-general}) gives an interaction independent of $y_0$, as required by translational invariance along the $y$ direction:
\begin{equation}
E_{\mathrm{int},\mathrm{cyc}}^{(s)}(x_0)
=
-
\frac{s\phi_0^2\alpha_0}{4\pi^2}
\frac{\sin\!\left(\frac{\pi x_0}{2w}\right)}
{1+\pi\lambda_{\mathrm{eff}}/w}.
\label{eq:eint-harmonic}
\end{equation}
As a consistency check, differentiating this interaction energy gives
\begin{equation}
F_{\mathrm{cyc},x}^{(s)}
=
-\frac{\partial E_{\mathrm{int},\mathrm{cyc}}^{(s)}}{\partial x_0}
=
\frac{s\phi_0}{c}j_{\mathrm{cyc},y}(x_0),
\label{eq:force-harmonic}
\end{equation}
which is the Lorentz force exerted by the screened sheet current on the vortex.

The periodic vortex potential can also favor the spontaneous formation of
vortex--antivortex pairs. To obtain a simple estimate, consider in zero
applied field a vortex of charge $s=+1$ at $x>0$ and an antivortex of
charge $s=-1$ at $-x$, placed on the corresponding attractive branches of
the periodic potential. For
$\xi\ll 2x\ll\lambda_{\mathrm{eff}}$, the energy of the pair in the
absence of the cycloidal texture is
\begin{equation}
E_{\mathrm{v\bar v}}^{(0)}(2x)
=
\frac{\phi_0^2}
{8\pi^2\lambda_{\mathrm{eff}}}
\ln\left(
\frac{2x}{\xi}
\right).
\label{eq:Epair-cyc-bare}
\end{equation}
Using Eq.~(\ref{eq:eint-harmonic}), the two vortices acquire the same
energy gain from the cycloidal potential. The corresponding total pair
energy is therefore
\begin{equation}
E_{\mathrm{pair,cyc}}(x)
=
\frac{\phi_0^2}
{8\pi^2\lambda_{\mathrm{eff}}}
\ln\left(
\frac{2x}{\xi}
\right)
-
\frac{\phi_0^2|\alpha_0|}
{2\pi^2}
\frac{
\sin\left(
\dfrac{\pi x}{2w}
\right)
}{
1+\pi\lambda_{\mathrm{eff}}/w
}.
\label{eq:Epair-cyc}
\end{equation}

A simple estimate of the threshold for spontaneous pair formation is
obtained by placing the vortex and antivortex close to the minima of their
respective cycloidal potentials, $x = \pm w$. Equation~(\ref{eq:Epair-cyc})
then gives
\begin{equation}
E_{\mathrm{pair,cyc}}(w)
\simeq
\frac{\phi_0^2}
{8\pi^2\lambda_{\mathrm{eff}}}
\ln\left(
\frac{2w}{\xi}
\right)
-
\frac{\phi_0^2|\alpha_0|}
{2\pi^2}
\frac{1}
{1+\pi\lambda_{\mathrm{eff}}/w}.
\label{eq:Epair-cyc-min-estimate}
\end{equation}

The condition
$E_{\mathrm{pair,cyc}}(w)<0$
therefore gives the estimate
\begin{equation}
|\alpha_0|\lambda_{\mathrm{eff}}
\gtrsim
\frac{1}{4}
\left(
1+\frac{\pi\lambda_{\mathrm{eff}}}{w}
\right)
\ln\left(
\frac{2w}{\xi}
\right).
\label{eq:alpha-c-cyc-estimate}
\end{equation}

Thus, already the periodic harmonic cycloid can in principle stabilize a
zero-field vortex--antivortex pair when the magnetoelectric coupling is
sufficiently strong. 

In known magnetic materials with cycloidal magnetic order -such as TbMnO$_3$, 
DyMnO$_3$, Ni$_3$V$_2$O$_8$, and BiFeO$_3$
\cite{Kenzelmann2005,Finger2014,Kenzelmann2006,Cabrera2009,
Bordacs2018,Meisenheimer2024}
- the period of magnetic moment rotation does not exceed 100 nm, making 
the criterion (\ref{eq:alpha-c-cyc-estimate}) for spontaneous vortex–antivortex 
generation difficult to achieve. In a weak external magnetic field, however, 
the proposed vortex pinning mechanism will decorate the magnetic structure with vortices, 
potentially giving rise to interesting commensurability phenomena between 
the periods of the magnetic system and the vortex lattice.

The estimate above neglects the displacement of the
vortex and antivortex from the minima of the single-vortex potential caused
by their mutual attraction. A more accurate threshold is obtained by
minimizing Eq.~(\ref{eq:Epair-cyc}) with respect to $x$. The corresponding
full minimization is carried out below for the isolated N\'eel wall.

The Fourier-space derivation of the Pearl kernel underlying
Eqs.~(\ref{eq:Ay-harmonic}) and~(\ref{eq:eint-harmonic}) is given in the Supplemental Material~\cite{SupplementalMaterial}.


The periodic harmonic cycloid therefore exhibits phase modulation, a
transverse magnetoelectric current, a sign-changing magnetic induction, and
a periodic vortex potential. This reference problem demonstrates the basic
electrodynamic consequences of a continuously rotating magnetoelectric
source in a geometry where the response is analytically transparent. We now
turn to the isolated N\'{e}el wall, for which the finite extent of the
rotation region produces a qualitatively different spatial structure of the
screened current, magnetic field, and vortex pinning potential.

\section{Equilibrium phase and spontaneous current of a single N\'{e}el wall}
\label{sec:model}

We now turn to the main geometry considered in this work: the single
N\'{e}el wall in the ferromagnetic layer introduced in
Sec.~\ref{sec:geometry}.

Below $T_c$, on length scales for which spatial variations of the order-parameter amplitude can be neglected, the electrodynamic response of the superconducting film is described within the London approximation. We therefore take the order-parameter amplitude to be uniform outside vortex cores and write $\psi(\mathbf{r})=\psi_0 e^{i\varphi(\mathbf{r})}$, 
with $\nabla\psi_0=0$. The superconducting response is then determined by the phase $\varphi(\mathbf{r})$ and the vector potential $\mathbf{A}$. Up to terms
independent of $\varphi$ and $\mathbf{A}$, the superconducting free-energy density can be written as 
\begin{equation}
f_{\mathrm{sc}}
=
\gamma\psi_0^2\,\left(\nabla\varphi-\frac{2e}{\hbar c}\mathbf{A}\right)^2
+2\gamma\psi_0^2\,\boldsymbol{\alpha}\cdot\left(\nabla\varphi-\frac{2e}{\hbar c}\mathbf{A}\right).
\label{eq:f_sc_gauge}
\end{equation}
The expression above differs from the completed-square form only by a term proportional to $\alpha^2(x)$, which is independent of $\varphi$ and $\mathbf{A}$ and therefore does not affect the variational equations.  Varying $F_{\mathrm{sc}}=d_s\int d^2r\,f_{\mathrm{sc}}$ with respect to
the vector potential, using
$\mathbf{j}_s(\mathbf{r}_\parallel)
=-c\,\delta F_{\mathrm{sc}}/
\delta\mathbf{A}(\mathbf{r}_\parallel)$,
gives the superconducting sheet current

\begin{equation}
\mathbf{j}_s(\mathbf{r}_\parallel)
=
d_s\frac{4e}{\hbar}\gamma\psi_0^2
\left(
\nabla\varphi-\frac{2e}{\hbar c}\mathbf{A}+\boldsymbol{\alpha}
\right).
\label{eq:j_gauge}
\end{equation}
The first two terms represent the conventional gauge-invariant condensate
momentum, whereas $\boldsymbol{\alpha}$ gives the anomalous
magnetoelectric contribution to the current. The corresponding unscreened
current, denoted by $\mathbf{j}_s^{(0)}$, is obtained by setting
$\mathbf{A}=0$.

For a one-dimensional texture, the phase-dependent part of
Eq.~(\ref{eq:f_sc_gauge}) at $\mathbf{A}=0$ is, up to an additive constant,
\begin{equation}
f_{\mathrm{phase}}
=
\gamma\psi_0^2
\left[
(\partial_x\varphi)^2
+2\alpha_x(x)\,\partial_x\varphi
\right].
\label{eq:f_phase_1d}
\end{equation}
The corresponding Euler--Lagrange equation is
\begin{equation}
\partial_x^2\varphi+\partial_x\alpha_x(x)=0.
\label{eq:euler_lagrange}
\end{equation}
This equation remains valid after screening is taken into account. Indeed, variation
of the full functional with respect to the phase gives
\begin{equation}
\nabla\cdot
\left(
\nabla\varphi-\frac{2e}{\hbar c}\mathbf{A}+\boldsymbol{\alpha}
\right)=0.
\label{eq:phase-london-full}
\end{equation}
For the one-dimensional geometry specified in
Sec.~\ref{sec:geometry}, with
$\mathbf{A}=A_y(x,z)\hat{\mathbf{y}}$,
Eq.~(\ref{eq:phase-london-full}) reduces to
$\partial_x(\partial_x\varphi+\alpha_x)=0$. The vanishing longitudinal
sheet current fixes the integration constant to zero, yielding
\begin{equation}
\partial_x\varphi(x)=-\alpha_x(x).
\label{eq:phase-gradient}
\end{equation}
Thus, the longitudinal component of the magnetoelectric source is fully compensated by the equilibrium phase gradient, and the phase profile obtained below is unaffected by transverse electromagnetic screening.

For the single N\'{e}el-wall source in Eq.~(\ref{eq:alpha-piecewise}), integration of Eq.~(\ref{eq:phase-gradient}) gives 
\begin{equation}
\varphi(x)=-\alpha_0
\begin{cases}
x+w, & x<-w,\\[4pt]
\dfrac{2w}{\pi}\cos\!\left(\frac{\pi x}{2w}\right), & |x|\le w,\\[10pt]
-(x-w), & x>w.
\end{cases}
\label{eq:phi-solution}
\end{equation}
Here, the arbitrary additive constant has been chosen such that
$\varphi(\pm w)=0$. Both the phase and its first derivative are continuous at $x=\pm w$. Outside the rotation region, for $x<-w$ and $x>w$, the phase gradient takes the constant values associated with the two helical states selected by the oppositely magnetized domains. In contrast, within the rotation region, it follows the spatial variation of the longitudinal magnetoelectric source.

The connection with the periodic cycloid considered in Sec.~\ref{sec:harmonic-london} is now transparent. Within the rotation region $|x|\leq w$, the longitudinal component $\alpha_x(x)$ is identical for the periodic cycloid and the single N\'{e}el wall. Consequently, the London phase profile in this region coincides with that obtained for the periodic texture. 

Setting $\mathbf{A}=0$ in Eq.~(\ref{eq:j_gauge}) and substituting
Eqs.~(\ref{eq:phi-solution}) and~(\ref{eq:alpha-piecewise}) gives the
unscreened sheet-current profile
\begin{equation}
j_{s,x}^{(0)}(x)=0,
\qquad
j_{s,y}^{(0)}(x)=j_0
\begin{cases}
0, & |x|>w,\\[4pt]
\cos\!\left(\frac{\pi x}{2w}\right), & |x|\le w.
\end{cases}
\label{eq:j_unscreened}
\end{equation}
Since the phase varies only along $x$, the $y$-component of $\boldsymbol{\alpha}$ is not compensated by the phase gradient and produces the transverse spontaneous current.

\section{Pearl screening and vortex pinning by a single N\'{e}el wall}
\label{sec:london}

In Sec.~\ref{sec:model}, we showed that the transverse component of the
magnetoelectric source generates a spontaneous current parallel to the wall. 

 The total free energy can be written as
\begin{equation}
F =
F_{0}
+
F_{\mathrm{film}}[\mathbf{A}]
+
F_{\mathrm{field}}[\mathbf{A}],
\end{equation}
where $F_{0}$ contains terms independent of the vector potential and
\begin{equation}
F_{\mathrm{field}}
=
\frac{1}{8\pi}
\int d^3r\,\mathbf{h}^2,
\end{equation}
is the magnetic-field energy. The film contribution is
\begin{equation}
F_{\mathrm{film}}
=
\frac{1}{8\pi\lambda_{\mathrm{eff}}}
\int d^2r\,
\left(
\frac{\phi_0}{2\pi}\nabla\varphi
+
\boldsymbol{\Phi}
-
\mathbf{A}
+
\frac{\phi_0}{2\pi}\boldsymbol{\alpha}
\right)^2.
\label{eq:film-free-energy}
\end{equation}
Here $\boldsymbol{\Phi}$ represents the vortex contribution to the
gauge-invariant condensate momentum. It is set to zero when calculating the
current and magnetic field generated by the N\'{e}el wall alone, but is
retained when the vortex--wall interaction is considered below.

Variation of $F_{\mathrm{film}}$ with respect to the vector potential gives
the superconducting sheet current
\begin{equation}
\mathbf{j}_s(\mathbf{r}_\parallel)
=
\frac{c}{4\pi\lambda_{\mathrm{eff}}}
\left[
\frac{\phi_0}{2\pi}\nabla\varphi
+\boldsymbol{\Phi}
-\mathbf{A}
+\frac{\phi_0}{2\pi}\boldsymbol{\alpha}
\right].
\label{eq:sheet-current}
\end{equation}
The corresponding three-dimensional current density is
$\mathbf{j}(\mathbf{r})
=
\mathbf{j}_s(\mathbf{r}_\parallel)\delta(z)$.

For the wall-only problem, $\boldsymbol{\Phi}=0$. Using the equilibrium
condition $\partial_x\varphi+\alpha_x=0$, derived in Sec.~\ref{sec:model}, the longitudinal contribution vanishes. The remaining transverse magnetoelectric source is
\begin{equation}
\mathbf{S}_{\mathrm{w}}(x)
=
\frac{\phi_0}{2\pi}
\alpha_y(x)\hat{\mathbf{y}},
\label{eq:wall-source}
\end{equation}
so that the sheet current can be written as
\begin{equation}
\mathbf{j}_s
=
\frac{c}{4\pi\lambda_{\mathrm{eff}}}
\left(
\mathbf{S}_{\mathrm{w}}-\mathbf{A}
\right).
\label{eq:sheet-current-wall}
\end{equation}
In the absence of electromagnetic screening,
$\mathbf{A}=0$, this expression reduces to the unscreened spontaneous current $\mathbf{j}_s^{(0)}$ obtained in Eq.~(\ref{eq:j_unscreened}).

In Coulomb gauge, Maxwell's equation takes the form
\begin{equation}
-\nabla^2\mathbf{A}(\mathbf{r})
=
\frac{4\pi}{c}
\mathbf{j}_s(\mathbf{r}_{\parallel})\delta(z).
\label{eq:maxwell-sheet}
\end{equation}
We use the two-dimensional Fourier convention
\begin{equation}
f_{\mathbf{k}}=\int d^2r\, f(\mathbf{r})e^{-i\mathbf{k}\cdot\mathbf{r}},
\qquad
f(\mathbf{r})=\int\frac{d^2k}{(2\pi)^2}\, f_{\mathbf{k}}e^{i\mathbf{k}\cdot\mathbf{r}}.
\label{eq:fourier-convention}
\end{equation}

For a Fourier component with $\mathbf{k}\neq0$, the solution that decays
away from the film has the form
\begin{equation}
\mathbf{A}_{\mathbf{k}}(z)
=
\mathbf{A}_{\mathbf{k}}(0)e^{-|\mathbf{k}||z|}.
\end{equation}
Integration of Eq.~(\ref{eq:maxwell-sheet}) across the film gives
\begin{equation}
\mathbf{A}_{\mathbf{k}}(0)
=
\frac{2\pi}{c|\mathbf{k}|}
\mathbf{j}_s(\mathbf{k}).
\label{eq:maxwell-sheet-boundary}
\end{equation}

In Fourier space, Eq.~(\ref{eq:sheet-current}) reads

\begin{equation}
\frac{4\pi\lambda_{\mathrm{eff}}}{c}
\mathbf{j}_s(\mathbf{k})
=
\frac{\phi_0}{2\pi}
\left[\nabla\varphi\right]_{\mathbf{k}}
+
\boldsymbol{\Phi}_{\mathbf{k}}
-
\mathbf{A}_{\mathbf{k}}(0)
+
\frac{\phi_0}{2\pi}\boldsymbol{\alpha}_{\mathbf{k}}.
\label{eq:sheet-current-fourier}
\end{equation}

Combining Eqs.~(\ref{eq:sheet-current-fourier}) and
(\ref{eq:maxwell-sheet-boundary}) yields the thin-film result
\begin{equation}
\mathbf{A}_{\mathbf{k}}(0)
=
\frac{
\displaystyle
\frac{\phi_0}{2\pi}
\left[\nabla\varphi\right]_{\mathbf{k}}
+\boldsymbol{\Phi}_{\mathbf{k}}
+\frac{\phi_0}{2\pi}\boldsymbol{\alpha}_{\mathbf{k}}
}{
1+2|\mathbf{k}|\lambda_{\mathrm{eff}}
}.
\label{eq:Aq}
\end{equation}
The denominator $1+2|\mathbf{k}|\lambda_{\mathrm{eff}}$
is the Pearl screening kernel. It determines how each spatial Fourier component of the magnetoelectric source contributes to the screened current and magnetic field.

For the single N\'{e}el wall, translational invariance along $y$
restricts the relevant Fourier components to $k_y=0$, so that
$\mathbf{k}=k_x\hat{\mathbf{x}}$. After the longitudinal source has been
compensated by the equilibrium phase gradient, only the transverse component
remains. The wall-induced vector potential is therefore
\begin{equation}
A_{k,y}(0)
=
\frac{
\displaystyle
\frac{\phi_0}{2\pi}\alpha_{k,y}
}{
1+2|k|\lambda_{\mathrm{eff}}
},
\label{eq:A-wall}
\end{equation}
which provides the starting point for the screened current and magnetic field derived below.

Equation~(\ref{eq:A-wall}) is written for the spin--orbit magnetoelectric
source alone, with $A_{k,y}(0)$ denoting the vector potential generated
self-consistently by the superconducting sheet current. A prescribed direct orbital contribution from the magnetic stray field of
the same single N\'{e}el wall enters the transverse London combination.
It can therefore be included, mode by mode, by replacing
$(\phi_0/2\pi)\alpha_{k,y}$ with
$(\phi_0/2\pi)\alpha_{k,y}-A^{\mathrm M}_{k,y}(0^-)$.
The Pearl denominator remains unchanged because it originates from the
self-consistent electromagnetic response of the superconducting sheet
current. The screened current, the screening field generated by the superconducting
sheet current, and the vortex--wall interaction then follow from the same
Fourier-space kernel with this signed transverse source. In the SOC-dominated regime specified in
Sec.~\ref{sec:harmonic-london}, we retain only the magnetoelectric term.

\subsection{Fourier transform of the wall source}

Accordingly, the two-dimensional Fourier transform of the
magnetoelectric source field takes the form
\begin{equation}
\boldsymbol{\alpha}_{\mathbf{k}}=(2\pi)\delta(k_y)\,\boldsymbol{\alpha}_{k_x},
\qquad
\boldsymbol{\alpha}_{k_x}=\int_{-\infty}^{\infty}dx\,\boldsymbol{\alpha}(x)e^{-ik_x x}.
\label{eq:alpha-q}
\end{equation}
The factor $\delta(k_y)$ expresses momentum conservation along the wall and reduces the subsequent vortex--wall interaction to an integral over $k_x$.

We first consider the contribution from the magnetization-rotation region,
$|x|\leq w$. For the N\'{e}el-wall profile in
Eq.~(\ref{eq:alpha-piecewise}), the Fourier components of the
magnetoelectric source in this region are
\begin{equation}
\begin{aligned}
\alpha_{\mathrm{rot},x}(k_x)
&=
2i\alpha_0k_x
\frac{\cos(k_xw)}
{\left(\frac{\pi}{2w}\right)^2-k_x^2},
\\
\alpha_{\mathrm{rot},y}(k_x)
&=
\frac{\alpha_0\pi}{w}
\frac{\cos(k_xw)}
{\left(\frac{\pi}{2w}\right)^2-k_x^2}.
\end{aligned}
\label{eq:alpha-rot-components}
\end{equation}

The apparent singularities at $k_x=\pm\pi/(2w)$ are removable, since both the numerator and the denominator vanish at these wave vectors. 

The homogeneous domains contribute only to the longitudinal component,
\begin{equation}
\boldsymbol{\alpha}_{\mathrm{hom}}(k_x)
=
2i\alpha_0\,\mathrm{P.V.}\!\left(\frac{\cos(k_x w)}{k_x}\right)\hat{\mathbf{x}},
\label{eq:alpha-uni-q}
\end{equation}
where $\mathrm{P.V.}$ denotes the Cauchy principal value. The complete Fourier transform of the longitudinal source is therefore 
\begin{equation}
\alpha_{k_x,x}
=
\alpha_{\mathrm{rot},x}(k_x)
+
\alpha_{\mathrm{hom},x}(k_x),
\end{equation}
while the transverse source is entirely confined to the rotation region,
\begin{equation}
\alpha_{k_x,y}
=
\alpha_{\mathrm{rot},y}(k_x).
\label{eq:transverse-wall-source}
\end{equation}

The phase gradient is $\nabla\varphi(x)=(\partial_x\varphi(x),0)$, where, for the single-wall
profile,
\begin{equation}
\partial_x\varphi(x)=\alpha_0
\begin{cases}
-1, & x<-w,\\[6pt]
\sin\!\left(\frac{\pi x}{2w}\right), & |x|\le w,\\[6pt]
+1, & x>w.
\end{cases}
\label{eq:dphi-piecewise}
\end{equation}

\subsection{Pearl-screened spontaneous current}

\begin{figure*}[t]
\centering

\begin{minipage}[t]{0.498\textwidth}
    \centering
    \includegraphics[width=\linewidth,trim=0 0 0 0,clip]{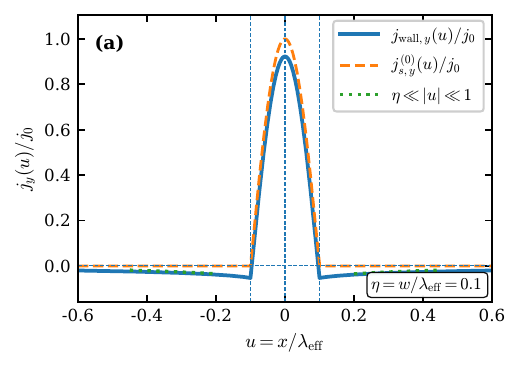}
\end{minipage}\hfill
\begin{minipage}[t]{0.498\textwidth}
    \centering
    \includegraphics[width=\linewidth,trim=0 0 0 0,clip]{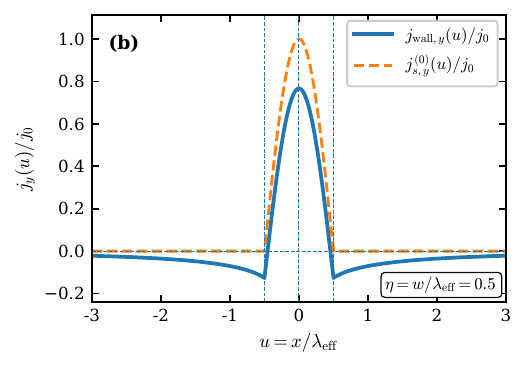}
\end{minipage}

\vspace{0.20cm}

\begin{minipage}[t]{0.498\textwidth}
    \centering
    \includegraphics[width=\linewidth,trim=0 0 0 0,clip]{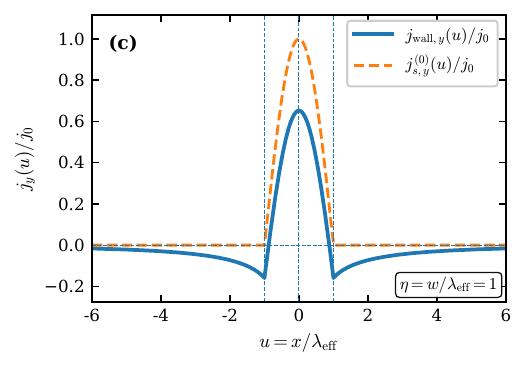}
\end{minipage}\hfill
\begin{minipage}[t]{0.498\textwidth}
    \centering
    \includegraphics[width=\linewidth,trim=0 0 0 0,clip]{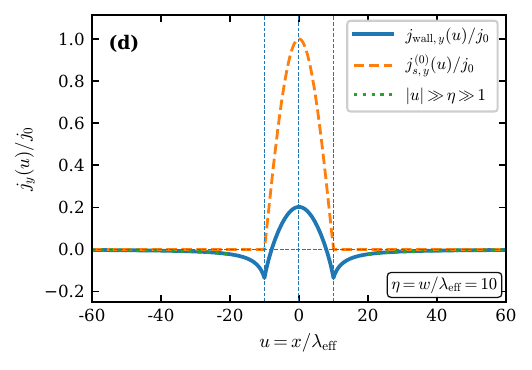}
\end{minipage}

\caption{Pearl-screened and unscreened spontaneous sheet-current profiles induced by the N\'{e}el wall. Solid curves show the Pearl-screened current $j_{\mathrm{wall},y}(u)/j_0$ obtained from Eq.~(\ref{eq:jwall-exact}), whereas dashed curves show the unscreened sheet current $j_{s,y}^{(0)}(u)/j_0$ from Eq.~(\ref{eq:j_unscreened}). Panels (a)--(d) correspond to $\eta=0.1$, $0.5$, $1.0$, and $10.0$, respectively.   The vertical dashed lines at $u=\pm\eta$ mark the boundaries
of the magnetization rotation region. The plotting range in each panel is
chosen to resolve both the current within this region and its long-range extension produced by Pearl screening. Dotted curves, where shown, represent the corresponding controlled asymptotic limits of the screened current rather than fits: panel (a) shows the logarithmic behavior in the intermediate range $\eta\ll |u|\ll1$ from Eq.~(\ref{eq:j-asymp-log}), while panel (d) shows the broad-wall far-field asymptote for $|u|\gg\eta\gg1$ given by Eq.~(\ref{eq:j-asymp-smallLambda}). Panels (b) and (c) correspond to the crossover between the narrow- and broad-wall regimes and are therefore shown without asymptotic curves.}
\label{fig:current}
\end{figure*}

Electromagnetic screening modifies the local current profile derived in
Eq.~(\ref{eq:j_unscreened}). Combining Eqs.~(\ref{eq:sheet-current}) and
(\ref{eq:Aq}), with $\boldsymbol{\Phi}=0$, gives the screened current in
Fourier space,
\begin{equation}
j_{s,y}(k_x)
=
j_{s,y}^{(0)}(k_x)
\frac{2|k_x|\lambda_{\mathrm{eff}}}
{1+2|k_x|\lambda_{\mathrm{eff}}},
\label{eq:screened-current-fourier}
\end{equation}
where $j_{s,y}^{(0)}(k_x)$ is the Fourier transform of the unscreened spontaneous current in Eq.~(\ref{eq:j_unscreened}). The factor $2|k_x|\lambda_{\mathrm{eff}}/[1+2|k_x|\lambda_{\mathrm{eff}}]$ is the Pearl screening factor for the sheet current. In particular, it vanishes at $k_x=0$, implying
\begin{equation}
\int_{-\infty}^{\infty}dx\,j_{s,y}(x)=0.
\label{eq:zero-net-current}
\end{equation}
Thus, although the unscreened current has a nonzero integral over the
magnetization rotation region, the self-consistent screened current carries
no net transport current. The compensating contribution is distributed
outside the rotation region through the long-range Pearl response.

Introducing the dimensionless coordinate $u=x/\lambda_{\mathrm{eff}}$ and
the dimensionless wall half-width $\eta=w/\lambda_{\mathrm{eff}}$, the
inverse Fourier transform gives
\begin{equation}
\begin{aligned}
j_{\mathrm{wall},y}(u) &= \frac{j_0}{\eta}
\Big[J_c(|u+\eta|)+J_c(|u-\eta|)\Big],
\end{aligned}
\label{eq:jwall-exact}
\end{equation}
with the same current scale $j_0$ as in
Eqs.~(\ref{eq:leading-current-harmonic}) and~(\ref{eq:j_unscreened}).
Equivalently, $j_0=c\phi_0\alpha_0/(8\pi^2\lambda_{\mathrm{eff}})$.
The kernel is
\begin{equation}
\begin{aligned}
J_c(\beta)
&= \frac{1}{\left(\dfrac{\pi}{\eta}\right)^2-1}\Bigg[
\mathcal{F}_0\!\left(\dfrac{\beta}{2}\right)
-\mathcal{F}_0\!\left(\dfrac{\pi\beta}{2\eta}\right)\\
&\qquad+\dfrac{\pi}{2}\left(\dfrac{\pi}{\eta}-1\right)
\sin\!\left(\dfrac{\pi\beta}{2\eta}\right)
\Bigg],
\end{aligned}
\label{eq:jc-exact}
\end{equation}
with
\begin{equation}
\mathcal{F}_0(t)=\cos(t)\,\mathrm{Ci}(t)+\sin(t)\left[\mathrm{Si}(t)-\frac{\pi}{2}\right].
\label{eq:fc-def}
\end{equation}
Here $\mathrm{Si}(t)=\int_0^t ds\,\sin(s)/s$ and
$\mathrm{Ci}(t)=-\int_t^\infty ds\,\cos(s)/s$ denote the sine and cosine integrals, respectively.
The apparent singularity at $\eta=\pi$ in
Eq.~(\ref{eq:jc-exact}) is removable, and the kernel at this point is
defined by the corresponding limiting value. Likewise, the limit
$\beta\to0$ is finite because the logarithmic singularities of the two
$\mathcal{F}_0$ terms cancel. The same limiting prescription applies below
to the magnetic-field and vortex--wall interaction kernels containing the
same denominator. Since $J_c$ depends on $|u\pm\eta|$, the current profile is
even,
\begin{equation}
j_{\mathrm{wall},y}(-u)
=
j_{\mathrm{wall},y}(u),
\end{equation}
as required by the reflection symmetry of the N\'{e}el-wall texture.

The effect of Pearl screening is illustrated in Fig.~\ref{fig:current}.
Before screening is included, the spontaneous current is confined to the
magnetization rotation region and follows
$\cos(\pi u/2\eta)$ for $|u|\leq\eta$. The screened current retains the
same reflection symmetry but extends beyond the rotation region and develops
long-range tails. Their spatial form depends on the ratio
$w/\lambda_{\mathrm{eff}}$.

For a narrow wall, $\eta\ll1$, the far-field behavior at $|u|\gg1$, corresponding to $w\ll\lambda_{\mathrm{eff}}$ and $|x|\gg\lambda_{\mathrm{eff}}$, is
\begin{equation}
\frac{j_{\mathrm{wall},y}(u)}{j_0}
\simeq
-\frac{8}{\pi^2}\frac{\eta}{u^2}
\left(
1-\frac{24}{u^2}
\right).
\label{eq:j-asymp-largeLambda}
\end{equation}
For a broad wall, $\eta\gg1$, the asymptotic profile at $|u|\gg\eta$, corresponding to $w\gg\lambda_{\mathrm{eff}}$ and $|x|\gg w$, becomes
\begin{equation}
\frac{j_{\mathrm{wall},y}(u)}{j_0}
\simeq
-\frac{8}{\pi^2}\frac{\eta}{u^2}
\left[
1
-
\frac{3(8-\pi^2)}{\pi^2}
\frac{\eta^2}{u^2}
\right].
\label{eq:j-asymp-smallLambda}
\end{equation}

For a narrow wall, there is also an intermediate spatial range 
$\eta\ll|u|\ll1$, or equivalently $w\ll|x|\ll\lambda_{\mathrm{eff}}$, 
in which the current varies logarithmically:

\begin{equation}
\frac{j_{\mathrm{wall},y}(u)}{j_0}
\simeq
\frac{2\eta}{\pi^2}\left(\gamma_E+\ln\dfrac{|u|}{2}\right),
\label{eq:j-asymp-log}
\end{equation}
where $\gamma_E$ is the Euler--Mascheroni constant. 

Figure~\ref{fig:current} compares the exact current profile with the
asymptotic expressions in the spatial ranges where the corresponding scale
separation is satisfied. For narrow walls, the near-wall behavior is
described by the logarithmic form in
Eq.~(\ref{eq:j-asymp-log}), whereas the algebraic asymptote in
Eq.~(\ref{eq:j-asymp-largeLambda}) is reached only at distances
$|x|\gg\lambda_{\mathrm{eff}}$. The cases $\eta=0.5$ and $\eta=1$
represent the crossover between the narrow- and broad-wall regimes, while
for $\eta=10$ the broad-wall asymptote in
Eq.~(\ref{eq:j-asymp-smallLambda}) is reached at distances large compared
with $w$.

The piecewise N\'eel-wall parametrization used above is continuous but has
derivative discontinuities at $x=\pm w$. These discontinuities produce an
algebraic large-$|k_x|$ decay of the corresponding Fourier transform and
can therefore affect the detailed current distribution on the scale of the
wall width. They are, however, not responsible for the long-distance
algebraic current tail. To verify this, we repeated the calculation for the
smooth micromagnetic profile
$\sin\theta=\tanh(x/2w)$ and
$\cos\theta=\operatorname{sech}(x/2w)$.
For this profile, the transverse source has the Fourier transform
\begin{equation}
\alpha_y(k_x)
=
2\pi w\alpha_0
\operatorname{sech}(\pi wk_x),
\label{eq:alpha-smooth-fourier-main}
\end{equation}
which decreases exponentially at large $|k_x|$. Nevertheless, Pearl
screening introduces the factor
$2\lambda_{\mathrm{eff}}|k_x|/
[1+2\lambda_{\mathrm{eff}}|k_x|]$.
Its nonanalytic small-$|k_x|$ behavior produces the same
$j_y(x)\propto-x^{-2}$ far-field dependence at
$|x|\gg w,\lambda_{\mathrm{eff}}$. Thus, the algebraic long-distance tail
is a consequence of nonlocal Pearl electrodynamics rather than of the
derivative discontinuities of the piecewise wall. The detailed derivation
and comparison of the two wall profiles are given in the Supplemental
Material.

\subsection{Magnetic field}

\begin{figure*}[t]
\centering

\begin{minipage}[t]{0.498\textwidth}
    \centering
    \includegraphics[width=\linewidth,trim=0 0 0 0,clip]{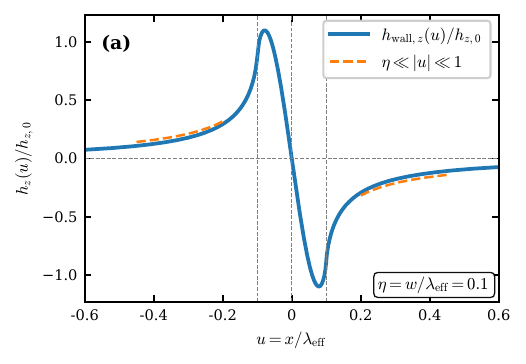}
\end{minipage}\hfill
\begin{minipage}[t]{0.498\textwidth}
    \centering
    \includegraphics[width=\linewidth,trim=0 0 0 0,clip]{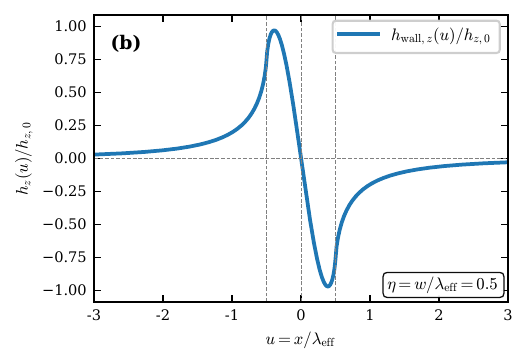}
\end{minipage}

\vspace{0.20cm}

\begin{minipage}[t]{0.498\textwidth}
    \centering
    \includegraphics[width=\linewidth,trim=0 0 0 0,clip]{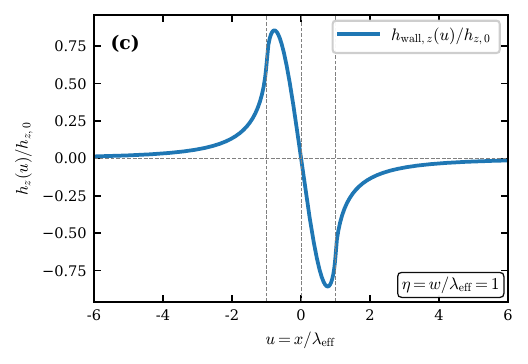}
\end{minipage}\hfill
\begin{minipage}[t]{0.498\textwidth}
    \centering
    \includegraphics[width=\linewidth,trim=0 0 0 0,clip]{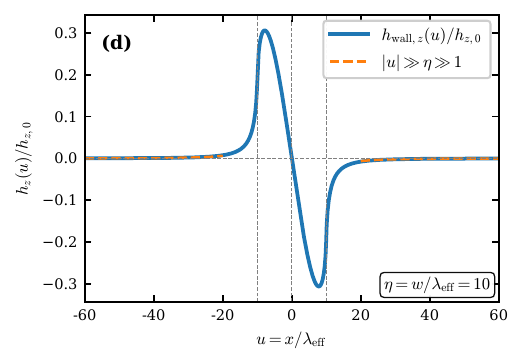}
\end{minipage}
\caption{Out-of-plane magnetic induction generated by the N\'{e}el wall. Solid curves show the exact field $h_{\mathrm{wall},z}(u)/h_{z,0}$ obtained from Eq.~(\ref{eq:hz-exact}). Panels (a)--(d) correspond to $\eta=0.1$, $0.5$, $1.0$, and $10.0$, respectively. The vertical dashed lines at $u=\pm\eta$ mark the boundaries of the magnetization rotation
region. The plotting range in each panel is chosen to resolve the field
reversal at the wall center and its spatial extension beyond the rotation
region. 
Dashed curves, when present, are controlled asymptotic limits of the exact field rather than fits: panel (a) shows the intermediate narrow-wall behavior for $\eta\ll |u|\ll1$ given by Eq.~(\ref{eq:h-asymp-intermediate}), while panel (d) shows the broad-wall far-field asymptote for $|u|\gg\eta\gg1$ given by Eq.~(\ref{eq:h-asymp-largeLambda}). Panels (b) and (c) correspond to the crossover between the narrow- and broad-wall regimes and are therefore shown without asymptotic curves.}
\label{fig:field}
\end{figure*}

The spontaneous current generated by the N\'{e}el wall produces a magnetic
field that extends into the surrounding three-dimensional space. For the one-dimensional wall geometry, the vector potential can be
written as
$\mathbf{A}(x,z)=A_y(x,z)\hat{\mathbf{y}}$.
The out-of-plane component of the magnetic induction at the superconducting
film is therefore $h_z(x)=\partial_x A_y(x,0)$, or, equivalently, in Fourier space,
$ h_z(k_x)=ik_x A_y(k_x,0)$.

After compensation of the longitudinal magnetoelectric source by the
equilibrium phase gradient, the vector potential is determined entirely by
the transverse component $\alpha_y(x)$. Substituting its Fourier transform
from Eq.~(\ref{eq:transverse-wall-source}) into
Eq.~(\ref{eq:A-wall}) and performing the inverse transform gives
the exact field profile in terms of the dimensionless coordinate:
\begin{equation}
\begin{aligned}
h_{\mathrm{wall},z}(u)
&=
\frac{\pi h_{z,0}}{2\eta}
\Big[
J_h(u+\eta)+J_h(u-\eta)
\Big],
\end{aligned}
\label{eq:hz-exact}
\end{equation}
where $h_{z,0} = \frac{\phi_0}{2\pi^2\lambda_{\mathrm{eff}}}\alpha_0$ and
\begin{equation}
\begin{aligned}
J_h(\beta) &= \operatorname{sgn}(\beta)\,
\frac{1}{\left(\dfrac{\pi}{\eta}\right)^2-1}\Bigg[
\mathcal{F}\!\left(\dfrac{|\beta|}{2}\right)
-\dfrac{\pi}{\eta}\mathcal{F}\!\left(\dfrac{\pi |\beta|}{2\eta}\right)\\
&\qquad+\dfrac{\pi}{2}\left(\dfrac{\pi}{\eta}-1\right)
\cos\!\left(\dfrac{\pi |\beta|}{2\eta}\right)
\Bigg].
\end{aligned}
\label{eq:jh-exact}
\end{equation}
Here the auxiliary function entering Eq.~(\ref{eq:jh-exact}) is
\begin{equation}
\mathcal{F}(t)=\sin(t)\,\mathrm{Ci}(t)-\cos(t)\left[\mathrm{Si}(t)-\frac{\pi}{2}\right].
\label{eq:fh-def}
\end{equation}
The explicit factor
$\operatorname{sgn}(\beta)$ makes $J_h(\beta)$ an odd function. Consequently,
\begin{equation}
h_{\mathrm{wall},z}(-u)
=
-h_{\mathrm{wall},z}(u),
\end{equation}
so that the magnetic field reverses sign across the center of the
N\'{e}el wall.
Figure~\ref{fig:field} shows the resulting field profiles for different
values of $\eta=w/\lambda_{\mathrm{eff}}$. Pearl screening produces a
magnetic field that extends well beyond the magnetization rotation region.
For a narrow wall, $\eta\ll1$, the field has an intermediate asymptotic
regime,

\begin{equation}
\frac{h_{\mathrm{wall},z}(u)}{h_{z,0}}
\simeq
-\frac{2\eta}{\pi}
\frac{\operatorname{sgn}(u)}{|u|},
\label{eq:h-asymp-intermediate}
\end{equation}
at $\eta \ll |u| \ll 1$, corresponding to $w\ll|x|\ll\lambda_{\mathrm{eff}}$.
Thus, between the wall scale and the Pearl screening length, the magnetic field decays as $1/|x|$.

At distances larger than both characteristic lengths, the field crosses over
to the algebraic far-field behavior
\begin{equation}
\frac{h_{\mathrm{wall},z}(u)}{h_{z,0}}
\simeq
-\frac{16\eta}{\pi}
\frac{\operatorname{sgn}(u)}{|u|^3}.
\label{eq:h-asymp-largeLambda}
\end{equation}

For a narrow wall, this asymptote applies when
$|u|\gg1\gg\eta$, corresponding to $|x|\gg\lambda_{\mathrm{eff}}\gg w$,
whereas for a broad wall it applies when
$|u|\gg\eta\gg1$, corresponding to $|x|\gg w\gg\lambda_{\mathrm{eff}}$.
The leading far-field decay is therefore the same in both limits,
$h_{\mathrm{wall},z}(x)\propto-\operatorname{sgn}(x)/|x|^3$, 
although the distance at which this asymptotic regime is reached depends on
the relative magnitude of $w$ and $\lambda_{\mathrm{eff}}$.

The exact profiles and the corresponding asymptotic expressions are compared
in Fig.~\ref{fig:field}. For $\eta=0.1$, the displayed range resolves the
intermediate $1/|u|$ behavior of
Eq.~(\ref{eq:h-asymp-intermediate}), while the ultimate $1/|u|^3$ decay is
reached only for $|u|\gg1$. For $\eta=10$, distances large compared with the
wall width approach the far-field behavior in
Eq.~(\ref{eq:h-asymp-largeLambda}). The cases $\eta=0.5$ and $\eta=1$
describe the crossover between these two limiting regimes.

The current and magnetic field have different spatial symmetries. The Pearl-screened current is even
under reflection about the wall center, whereas the magnetic field is odd.
The magnetic-field reversal at the wall center is therefore a direct
consequence of the reflection symmetry of the wall-induced current
distribution.

\subsection{Vortex--wall interaction potential}

\begin{figure*}[t]
\centering

\begin{minipage}[t]{0.498\textwidth}
    \centering
    \includegraphics[width=\linewidth,trim=0 0 0 0,clip]{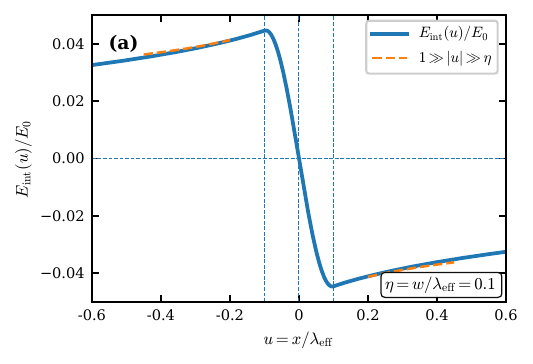}
\end{minipage}\hfill
\begin{minipage}[t]{0.498\textwidth}
    \centering
    \includegraphics[width=\linewidth,trim=0 0 0 0,clip]{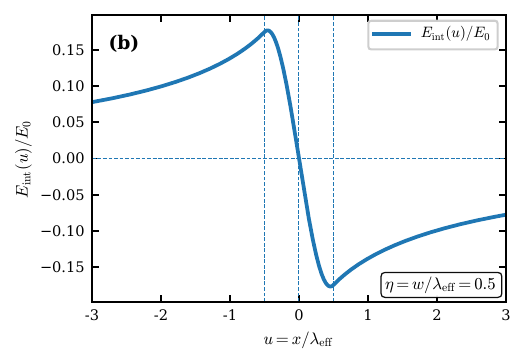}
\end{minipage}

\vspace{0.20cm}

\begin{minipage}[t]{0.498\textwidth}
    \centering
    \includegraphics[width=\linewidth,trim=0 0 0 0,clip]{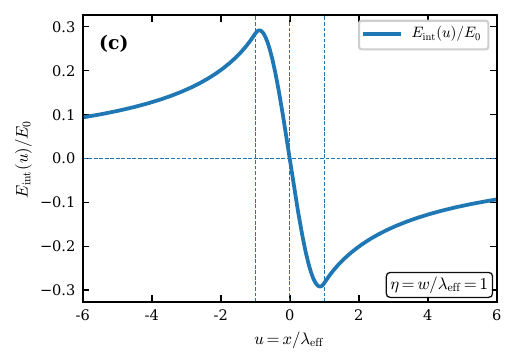}
\end{minipage}\hfill
\begin{minipage}[t]{0.498\textwidth}
    \centering
    \includegraphics[width=\linewidth,trim=0 0 0 0,clip]{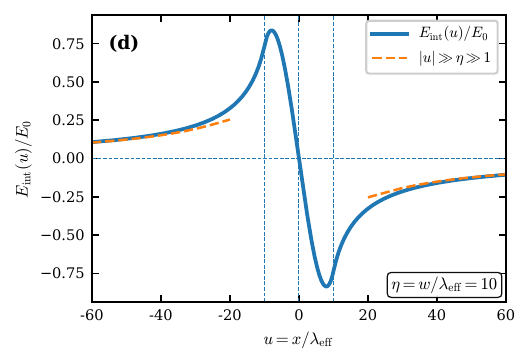}
\end{minipage}
\caption{Vortex--wall interaction potential for a fixed topological charge. Solid curves show the exact Pearl-screened interaction $E_{\mathrm{int}}(u)/E_0$ obtained from Eq.~(\ref{eq:eint-exact}). Panels (a)--(d) correspond to
$\eta=0.1$, $0.5$, $1.0$, and $10.0$, respectively. The vertical dashed
lines at $u=\pm\eta$ mark the boundaries of the magnetization rotation
region, and the line at $u=0$ marks the wall center. For the sign convention
used in Eq.~(\ref{eq:eint-exact}), a vortex is attracted to the $u>0$ side
of the wall and repelled from the $u<0$ side; the interaction is reversed for
an antivortex. Dashed curves, where shown, represent controlled asymptotic
expressions rather than fits: panel (a) shows the intermediate narrow-wall behavior for $1\gg |u|\gg\eta$ given by Eq.~(\ref{eq:eint-asymp-log}), while panel (d) shows the broad-wall far-field asymptote for $|u|\gg\eta\gg1$ given by Eq.~(\ref{eq:eint-asymp-smallLambda}). Panels (b) and (c) correspond to the crossover between the narrow- and broad-wall regimes and are therefore shown without asymptotic curves.}
\label{fig:interaction}
\end{figure*}

The magnetic field and spontaneous current generated by the N\'{e}el wall
also modify the energy of a vortex in the superconducting film. We now
derive the resulting vortex--wall interaction potential within the same
Pearl-screened London description used above. It is convenient to write the contribution of the equilibrium wall texture to the gauge-invariant condensate momentum as $\mathbf{C}=(\phi_0/2\pi)(\nabla\varphi+\boldsymbol{\alpha})$ and the singular contribution of a vortex with topological charge $s=\pm1$ as $\boldsymbol{\Phi}^{(s)}=(\phi_0/2\pi)\nabla\varphi_{\mathrm{vor}}^{(s)}$. Here $s=+1$ corresponds to a vortex and $s=-1$ to an antivortex. The film free energy then depends on the combination $
\mathbf{C} + \boldsymbol{\Phi}^{(s)} - \mathbf{A}$. After the vector potential is eliminated, the part of the free energy bilinear in the wall and vortex contributions defines their interaction
energy.

For a general Fourier component, the vector potential generated by the
combined wall and vortex sources is
\begin{equation}
\mathbf{A}_{\mathbf{k}}(0)
=
\frac{
\mathbf{C}_{\mathbf{k}}
+
\boldsymbol{\Phi}_{\mathbf{k}}^{(s)}
}{
1+2|\mathbf{k}|\lambda_{\mathrm{eff}}
}.
\label{eq:A-wall-vortex}
\end{equation}
The Pearl denominator is identical to that obtained for the wall-induced
current and magnetic field. The corresponding vector potential away from
the film is
\begin{equation}
\mathbf{A}_{\mathbf{k}}(z)
=
\mathbf{A}_{\mathbf{k}}(0)
e^{-|\mathbf{k}||z|}.
\end{equation}
Equivalently, after Fourier transformation with respect to $z$,
\begin{equation}
\mathbf{A}_{\mathbf{k},k_z}
=
\frac{2|\mathbf{k}|}
{k_z^2+|\mathbf{k}|^2}
\mathbf{A}_{\mathbf{k}}(0).
\end{equation}

The equilibrium phase gradient compensates the longitudinal component of the
magnetoelectric source, $\partial_x\varphi+\alpha_x=0$, so that the wall contribution $\mathbf{C}$ is purely transverse. The vortex
contribution is also transverse, $\mathbf{k}\cdot \boldsymbol{\Phi}_{\mathbf{k}}^{(s)}
= 0 $. The Coulomb-gauge solution therefore involves only transverse Fourier
components.
 
Substitution of the screened vector potential into the film and
magnetic-field energies gives the vortex--wall interaction

\begin{equation}
\begin{aligned}
E_{\mathrm{int}}^{(s)}
={}&
\frac{1}{4\pi}
\int\frac{d^2k}{(2\pi)^2}
\frac{|\mathbf{k}|}
{1+2|\mathbf{k}|\lambda_{\mathrm{eff}}}
\\
&\times
\left[
\mathbf{C}_{\mathbf{k}}
\cdot
\boldsymbol{\Phi}_{\mathbf{k}}^{(s)*}
+
\mathbf{C}_{\mathbf{k}}^{*}
\cdot
\boldsymbol{\Phi}_{\mathbf{k}}^{(s)}
\right].
\end{aligned}
\label{eq:eint-general}
\end{equation}

The factor
$|\mathbf{k}|/(1+2|\mathbf{k}|\lambda_{\mathrm{eff}})$
is the Pearl-screened interaction kernel. It determines how the spatial
Fourier components of the N\'{e}el-wall texture couple to the circulating
superflow of a vortex.

For the single N\'{e}el wall, only the transverse component of the
magnetoelectric source remains after the equilibrium phase has been taken
into account. Using the Fourier transform of
Eq.~(\ref{eq:alpha-q}), one obtains
\begin{equation}
\mathbf{C}_{\mathbf{k}}
=
\phi_0\delta(k_y)
\alpha_{k_x,y}\hat{\mathbf{y}},
\label{eq:C-wall-fourier}
\end{equation}
where $\alpha_{k_x,y}$ is given by
Eq.~(\ref{eq:transverse-wall-source}).

A vortex located at
$\mathbf{R}_0=(x_0,y_0)$ is represented in Fourier space by
\cite{deGennes1966,Pearl1964,Olde2019}
\begin{equation}
\boldsymbol{\Phi}_{\mathbf{k}}^{(s)}
=
\frac{\phi_0}{2\pi}
\nabla\varphi_{\mathrm{vor}}^{(s)}(\mathbf{k})
e^{-i\mathbf{k}\cdot\mathbf{R}_0},
\label{eq:vortex-source-fourier}
\end{equation}
with
\begin{equation}
\nabla\varphi_{\mathrm{vor}}^{(s)}(\mathbf{k})
=
2\pi i s\,
\frac{
k_y\hat{\mathbf{x}}
-
k_x\hat{\mathbf{y}}
}{
k_x^2+k_y^2
}.
\label{eq:vortex-phase-fourier}
\end{equation}
The reality condition is  $ \boldsymbol{\Phi}_{-\mathbf{k}}^{(s)} = \boldsymbol{\Phi}_{\mathbf{k}}^{(s)*}$.

Because the wall source contains $\delta(k_y)$, the $k_y$ integration in
Eq.~(\ref{eq:eint-general}) sets $k_y=0$. The interaction is therefore
independent of the vortex coordinate $y_0$, as expected from translational
invariance along the wall. Substituting the transverse wall source gives 
\begin{equation}
\begin{aligned}
E_{\mathrm{int}}^{(s)}(x_0)
={}&
-s\frac{2\alpha_0\pi\phi_0^2}
{w(2\pi)^3}
\int_0^\infty dk_x
\\
&\times
\frac{
\sin(k_xx_0)\cos(k_xw)
}{
\left[
\left(\dfrac{\pi}{2w}\right)^2-k_x^2
\right]
\left(
1+2k_x\lambda_{\mathrm{eff}}
\right)
}.
\end{aligned}
\label{eq:e-int-cos}
\end{equation}
The apparent singularity at
$k_x=\pi/(2w)$ is removable because the zero of
$\cos(k_xw)$ cancels the zero of the denominator.

In the following, we take $s=+1$ and denote the vortex position simply by
$x$. The interaction for an antivortex has the opposite sign,
\begin{equation}
E_{\mathrm{int}}^{(-)}(x)
=
-E_{\mathrm{int}}^{(+)}(x).
\end{equation}
We also take $\alpha_0>0$ without loss of generality. Reversing the sign of
$\alpha_0$, or equivalently reversing the chirality of the magnetic texture,
interchanges the attractive and repulsive sides of the wall.

In the same dimensionless variables, and with the positive scale $E_0=\alpha_0\phi_0^2/(2\pi^3)$, one obtains
 \begin{equation}
E_{\mathrm{int}}(u)=
-\frac{E_0}{2}\Big[J_{\mathrm{int}}(u+\eta)+J_{\mathrm{int}}(u-\eta)\Big].
\label{eq:eint-exact}
\end{equation}
The function $J_{\mathrm{int}}$ is defined by
\begin{equation}
\begin{aligned}
J_{\mathrm{int}}(\beta) &= \operatorname{sgn}(\beta)\, 
\frac{1}{\left(\dfrac{\pi}{\eta}\right)^2-1}\Bigg[
\dfrac{\pi}{\eta}\,\mathcal{F}\!\left(\dfrac{|\beta|}{2}\right)
-\mathcal{F}\!\left(\dfrac{\pi |\beta|}{2\eta}\right)\\
&\qquad-\dfrac{\pi}{2}\left(\dfrac{\pi}{\eta}-1\right)
\cos\!\left(\dfrac{\pi |\beta|}{2\eta}\right)
\Bigg].
\end{aligned}
\label{eq:jint-exact}
\end{equation}
The function $\mathcal{F}$ is defined in Eq.~(\ref{eq:fh-def}). Since $J_{\mathrm{int}}$ is odd, $E_{\mathrm{int}}(-u)=-E_{\mathrm{int}}(u)$ for a fixed topological charge. The wall-induced interaction, therefore, attracts a vortex of a given vorticity on one side of the wall and repels the same vorticity on the other side; reversing the topological charge reverses the potential. The wall thus produces an antisymmetric pinning potential that spatially separates the energetically preferred positions of vortices and antivortices.

Figure~\ref{fig:interaction} shows the interaction potential for different
values of $\eta=w/\lambda_{\mathrm{eff}}$. For a narrow wall,
$\eta\ll1$, the intermediate range $\eta\ll|u|\ll1$
corresponding to $ w\ll|x|\ll\lambda_{\mathrm{eff}} $ is described by 
\begin{equation}
\begin{aligned}
\frac{E_{\mathrm{int}}(u)}{E_0}
\simeq
-\frac{\eta}{2}\,
\operatorname{sgn}(u)
\Bigg[
1
+
\frac{|u|}{\pi}
\left(
\gamma_E+\ln\frac{|u|}{2}-1
\right)
\Bigg].
\end{aligned}
\label{eq:eint-asymp-log}
\end{equation}
The interaction, therefore, approaches a sign-changing plateau on the two
sides of a narrow wall, with logarithmic spatial corrections.

At distances larger than both the wall width and the Pearl screening length,
the interaction has an algebraic $1/|x|$ tail. For
$|u|\gg1\gg\eta$, corresponding to $|x|\gg\lambda_{\mathrm{eff}}\gg w$, one finds
\begin{equation}
\frac{E_{\mathrm{int}}(u)}{E_0}
\simeq
-\frac{2\eta}{\pi}
\frac{\operatorname{sgn}(u)}{|u|}
\left(
1-\frac{8}{|u|^2}
\right).
\label{eq:eint-asymp-largeLambda}
\end{equation}
For a broad wall,
$|u|\gg\eta\gg1$, corresponding to $|x|\gg w\gg\lambda_{\mathrm{eff}}$, the asymptote is
\begin{equation}
\frac{E_{\mathrm{int}}(u)}{E_0}
\simeq
-\frac{2\eta}{\pi}
\frac{\operatorname{sgn}(u)}{|u|}
\left(
1-\frac{8\eta^2}{\pi^2|u|^2}
\right).
\label{eq:eint-asymp-smallLambda}
\end{equation}

The exact interaction profiles and the corresponding asymptotic expressions
are compared in Fig.~\ref{fig:interaction}. For $\eta=0.1$, the displayed
range resolves the intermediate sign-changing plateau described by
Eq.~(\ref{eq:eint-asymp-log}); the ultimate narrow-wall $1/|u|$ tail is
reached only for $|u|\gg1$. For $\eta=10$, distances large compared with
the wall width approach the broad-wall asymptote in
Eq.~(\ref{eq:eint-asymp-smallLambda}). The cases $\eta=0.5$ and $\eta=1$
describe the crossover between these limiting regimes.

A useful analytical measure of the interaction strength is obtained by
evaluating the potential at the boundary of the magnetization rotation
region, $x=w$, or equivalently $u=\eta$:

\begin{equation}
E_{\mathrm{int}}(w)
=
-\frac{\alpha_0\phi_0^2}{4\pi^3}
\left[
\frac{\dfrac{\pi}{\eta}\mathcal{F}(\eta)-\mathcal{F}(\pi)}{\left(\dfrac{\pi}{\eta}\right)^2-1}
+
\frac{\pi}{2\left(\dfrac{\pi}{\eta}+1\right)}
\right].
\label{eq:eint-edge-exact}
\end{equation}
This quantity provides a convenient reference scale for the wall-induced
vortex interaction. It should be distinguished from the exact pinning
energy, which is determined by the minimum of the full interaction potential in Eq.~(\ref{eq:eint-exact}).

For a narrow wall, $\eta\ll1$ ($w\ll\lambda_{\mathrm{eff}}$), the interaction at the boundary of the
rotation region becomes
\begin{equation}
E_{\mathrm{int}}(w)\simeq
-\frac{w}{\lambda_{\mathrm{eff}}}\frac{E_0}{2}.
\label{eq:eint-edge-asymp}
\end{equation}
This expression gives the leading wall-edge pinning scale in the narrow-wall regime, measuring the energetic gain of placing a vortex on the attractive branch near the wall boundary.

The magnitude of this scale is controlled by
$w/\lambda_{\mathrm{eff}}$. For the representative ratio
$\xi/\lambda_{\mathrm{eff}}\sim10^{-3}$ considered in
Sec.~\ref{sec:geometry}, Eq.~(\ref{eq:eint-edge-asymp}) gives
$E_{\mathrm{int}}(w)/E_0\simeq
-5.0\times10^{-4}(w/\xi)$.
The narrow-wall London regime,
$w\ll \lambda_{\mathrm{eff}}$, then corresponds to
$1\ll w/\xi\ll10^3$, or equivalently
$\xi\ll w\ll\lambda_{\mathrm{eff}}$.
The absolute interaction energy remains proportional to
the magnetoelectric coupling through $E_0\propto\alpha_0$.

In the opposite limit, $\eta\gg1$ ($w\gg\lambda_{\mathrm{eff}}$), the interaction at the boundary of the rotation region approaches
\begin{equation}
E_{\mathrm{int}}(w)\simeq
-\frac{E_0}{2}
\left[\frac{\pi}{2}+\mathcal{F}(\pi)\right],
\label{eq:eint-edge-widewall}
\end{equation}
so that the wall-edge pinning scale becomes essentially independent of $w$ at leading order. Since $\frac{\pi}{2}+\mathcal{F}(\pi)=\mathrm{Si}(\pi)\simeq1.85194$, the saturated interaction energy is $E_{\mathrm{int}}(w)\simeq -0.926\,E_0$. Taken together, these limits show that the N\'{e}el wall provides an extended magnetoelectric pinning source whose characteristic energy scale can reach values of order $E_0$ in the wide-wall regime.

The single N\'{e}el wall thus generates an extended, antisymmetric vortex
pinning potential. Its sign is reversed by changing either the vortex
topological charge or the sign of the magnetoelectric coupling, while its
spatial range is set by the combined effects of the wall profile and Pearl
screening. This interaction provides the basis for the vortex--antivortex
configurations considered in Sec.~\ref{sec:pairs}.
 
Fig.~\ref{fig:interaction} illustrates the resulting antisymmetric interaction profile for a fixed topological charge. The exact minimum is obtained from the full profile, while the wall-edge value provides a useful analytical pinning scale. The asymptotic forms in
Eqs.~(\ref{eq:eint-asymp-log})--(\ref{eq:eint-asymp-smallLambda})
show that the pinning landscape remains long-ranged, reflecting the
extended character of the magnetoelectric source.

The vortex--wall interaction profile in Fig.~\ref{fig:interaction}
provides the static asymmetry required for a vortex-ratchet response
\cite{Lee1999,Villegas2003,Wordenweber2004,deSouzaSilva2006}. For
the wall profile and sign convention adopted here, a vortex is attracted
to the $x>0$ side and repelled from the $x<0$ side, so that the wall
produces an oriented barrier--well potential; for an antivortex, the
orientation is reversed. A vortex initially trapped in the attractive
well encounters inequivalent escape paths: motion toward $+x$ requires
leaving the well, whereas motion toward $-x$ additionally requires
crossing the repulsive barrier on the opposite side of the wall.

A transport current applied along the wall exerts a Lorentz force
transverse to the wall and periodically tilts this barrier--well
landscape. A suitable zero-mean ac transport current may drive a vortex across the
lower barrier during one half-cycle while the vortex remains pinned during
the opposite half-cycle. An array of N\'{e}el walls that repeats the same barrier--well
orientation may therefore provide a ratchet landscape capable of
converting the ac drive into a net vortex displacement, provided that the superposed array potential remains noncentrosymmetric.

The attractive well of the wall-induced interaction potential also provides
a mechanism for vortex dragging by a moving N\'{e}el wall. As the wall
moves, the pinning minimum is translated together with the magnetic
texture. A vortex trapped in this minimum can remain bound to the moving
wall and consequently be transported with it, provided that it remains
pinned during the wall motion.

We considered the vortex--wall interaction in the limit of a dominant SO interaction, $\left|\alpha_0\right|\phi_0/2\pi \gg 2\pi m d_f$.
A similar situation is realized in the inverse case, in which the interaction is dominated by the magnetic induction of the N\'{e}el wall (see Sec.~\ref{sec:harmonic-london}). To describe this case at the level of the interaction magnitude, it suffices to replace $\left|\alpha_0\right|\phi_0/2\pi$ by $2\pi m d_f$ in the corresponding formulas.

\section{N\'{e}el-wall-induced stabilization of vortex--antivortex pairs}
\label{sec:pairs}

The antisymmetric vortex potential derived in Sec.~\ref{sec:london} favors
opposite topological charges on opposite sides of the N\'{e}el wall. We now
determine whether this magnetoelectric energy gain can overcome the energy
cost of creating a vortex--antivortex pair in the superconducting film.

In zero applied magnetic field and in the absence of externally imposed net
vorticity, the relevant configuration has zero total topological charge. We
therefore consider a symmetric vortex--antivortex pair, with a vortex at
$x>0$ and an antivortex at $-x$. Their separation is $2x$. The total pair
energy relative to the vortex-free state is
\begin{align}
E_{\mathrm{pair}} & =E_{\mathrm{vor}}+E_{\mathrm{av}}+E_{\mathrm{int}}^{%
\mathrm{vor}}\left( x\right) +E_{\mathrm{int}}^{\mathrm{av}}\left( -x\right)
+U_{\mathrm{v\bar{v}}}\left( 2x\right) \notag \\
& =E_{\mathrm{v\bar v}}^{(0)}(2x)+E_{\mathrm{int}}^{\mathrm{vor}}+E_{\mathrm{int}}^{%
\mathrm{av}}.  
\label{eq:Epair-total}
\end{align}

Here $E_{\mathrm{vor}}$ and $E_{\mathrm{av}}$ are the self-energies of an
isolated Pearl vortex and antivortex,
$U_{\mathrm{v\bar v}}(2x)$ is their mutual interaction, and the last two
terms describe the interaction of the vortex and antivortex with the
N\'{e}el-wall-induced magnetoelectric potential.

The vortex and antivortex have the same self-energy, 
\begin{equation}
E_{\mathrm{vor}}=E_{\mathrm{av}}=\frac{\phi_{0}^{2}}{16\pi^{2}\lambda _{%
\mathrm{eff}}}\ln\left( \frac{\lambda_{\mathrm{eff}}}{\xi}\right).
\label{main:S22cc}
\end{equation}
For separations in the logarithmic regime,
$\xi\ll2x\ll\lambda_{\mathrm{eff}}$, the attractive vortex--antivortex interaction is
\begin{equation}
U_{\mathrm{v\bar v}}(2x)
=
-\frac{\phi_0^2}
{8\pi^2\lambda_{\mathrm{eff}}}
\ln\left(
\frac{\lambda_{\mathrm{eff}}}{2x}
\right).
\label{eq:vav-interaction}
\end{equation}
The energy of the pair in the absence of the wall-induced magnetoelectric
potential is therefore
\begin{equation}
\begin{aligned}
E_{\mathrm{v\bar v}}^{(0)}(2x)
={}&
E_{\mathrm{vor}}
+
E_{\mathrm{av}}
+
U_{\mathrm{v\bar v}}(2x)
\\
={}&
\frac{\phi_0^2}
{8\pi^2\lambda_{\mathrm{eff}}}
\ln\left(
\frac{2x}{\xi}
\right).
\end{aligned}
\label{eq:Epair-no-wall}
\end{equation}
Thus, the mutual attraction reduces the energy cost from that of two
independent Pearl vortices to a logarithmic pair energy determined by their
separation $2x$.

For the N\'{e}el-wall texture considered here, the interaction potential is
antisymmetric in position and changes sign with the topological charge.
Consequently, a vortex and an antivortex placed symmetrically on their
respective attractive sides have equal interaction energies:
 
\begin{equation}
E_{\mathrm{int}}^{\mathrm{vor}}\left( x\right) =E_{\mathrm{int}}^{\mathrm{av}%
}\left( -x\right) \equiv E_{\mathrm{int}}. 
\label{main:S29cc}
\end{equation}%
where%
\begin{equation}
E_{\mathrm{int}}(x)=-\alpha _{0}\frac{\phi _{0}^{2}}{4\pi ^{3}}\left[ J_{%
\mathrm{int}}\!\left( \frac{x+w}{\lambda _{\mathrm{eff}}}\right) +J_{\mathrm{%
int}}\!\left( \frac{x-w}{\lambda _{\mathrm{eff}}}\right) \right].
\label{eq:Eint_pair_position}
\end{equation}%
The total pair energy is therefore 
\begin{equation}
E_{\mathrm{pair}}(x)
=
E_{\mathrm{v\bar v}}^{(0)}(2x)
+
2E_{\mathrm{int}}(x).
\label{eq:Epair-compact}
\end{equation}
The first term favors a smaller vortex--antivortex separation, whereas the
wall-induced potential favors placing the two topological charges on opposite
sides of the wall. The competition between these two contributions determines the equilibrium separation.

The optimal position of the vortex--antivortex pair is obtained from $dE_{%
\mathrm{pair}}/dx=0$. Within the logarithmic pair-interaction approximation, $\xi \ll 2x\ll
\lambda _{\mathrm{eff}}$, this condition becomes 
\begin{equation}
\frac{\alpha _{0}\lambda _{\mathrm{eff}}}{4\pi ^{3}}\left[ J_{\mathrm{int}%
}^{\prime }\left( u+\eta \right) +J_{\mathrm{int}}^{\prime }\left( u-\eta
\right) \right] =\frac{1}{16\pi ^{2}u}.  \label{eq:xstar_stationarity}
\end{equation}%

The right-hand side originates from the vortex--antivortex attraction,
whereas the left-hand side describes the force exerted by the
N\'{e}el-wall-induced magnetoelectric potential. Details of the wall-edge
calculation and pair-energy minimization, including an explicit expression
for $J_{\mathrm{int}}'(\beta)$ and complementary scans at fixed
$\alpha_0\lambda_{\mathrm{eff}}$, are given in the Supplemental
Material~\cite{SupplementalMaterial}.

For the numerical analysis, we use the dimensionless pair energy
\begin{equation}
\widetilde{E}_{\mathrm{pair}}(x)\equiv
\frac{8\pi ^{2}\lambda _{\mathrm{eff}}}{\phi _{0}^{2}}
E_{\mathrm{pair}}(x).
\label{eq:Epair_dimensionless}
\end{equation}

All lengths are expressed in units of $\lambda_{\mathrm{eff}}$. The calculations of the equilibrium position and pair energy use
$\xi/\lambda_{\mathrm{eff}}=10^{-3}$, while the critical-coupling analysis
below shows explicitly the dependence on this ratio.

\begin{figure}[tb]
\resizebox{0.98\columnwidth}{!}{\rotatebox{0}{%
\includegraphics{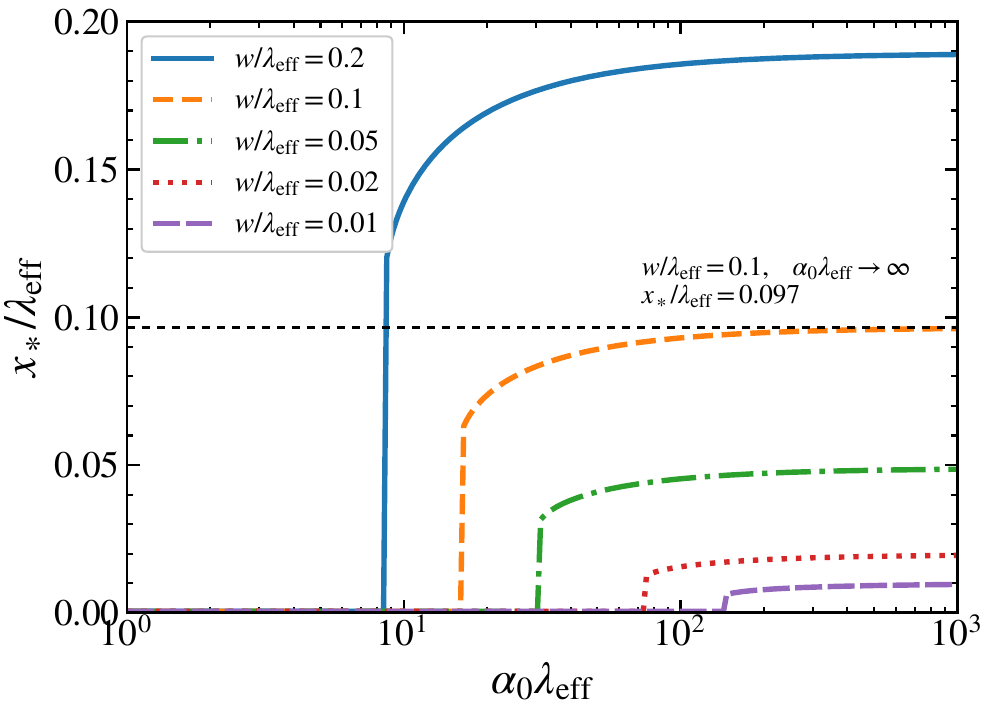}}}
\caption{Dependence of the optimal vortex--antivortex position $x_{\ast }/%
\protect\lambda _{\mathrm{eff}}$ on the dimensionless magnetoelectric
coupling strength $\protect\alpha_0\protect\lambda_{\mathrm{eff}}$ for
several values of the ratio $w/\protect\lambda _{\mathrm{eff}}$. }
\label{fig.07}
\end{figure}

Figure~\ref{fig.07} shows the equilibrium vortex position
$u_*=x_*/\lambda_{\mathrm{eff}}$ as a function of the dimensionless
magnetoelectric coupling
$\alpha_0\lambda_{\mathrm{eff}}$ for several values of
$\eta=w/\lambda_{\mathrm{eff}}$. The equilibrium position is obtained by
minimizing the total pair energy in Eq.~(\ref{eq:Epair-compact}) within the
range of validity of the logarithmic vortex--antivortex interaction,
\begin{equation}
\xi\ll2x_*\ll\lambda_{\mathrm{eff}}.
\end{equation}
The wall-induced potential tends to place the vortex and antivortex on
opposite sides of the wall, whereas their mutual attraction tends to reduce
the separation $2x_*$. Increasing the magnetoelectric coupling strengthens
the wall contribution, and the equilibrium positions approach the minima
selected by the isolated vortex--wall potential.

In the formal strong-coupling limit,
$\alpha_0\lambda_{\mathrm{eff}}\to\infty$, the contribution of the
vortex--antivortex attraction to the stationarity condition becomes
negligible relative to the wall-induced force. The equilibrium position then
approaches the minimum of the single-vortex interaction potential. For
$\eta=0.1$, this gives
\begin{equation}
\frac{x_*}{\lambda_{\mathrm{eff}}}
\simeq0.097.
\end{equation}
At finite coupling, the mutual vortex--antivortex attraction shifts the two defects toward the wall center and reduces their equilibrium separation.

\begin{figure}[tb]
\resizebox{0.98\columnwidth}{!}{\rotatebox{0}{%
\includegraphics{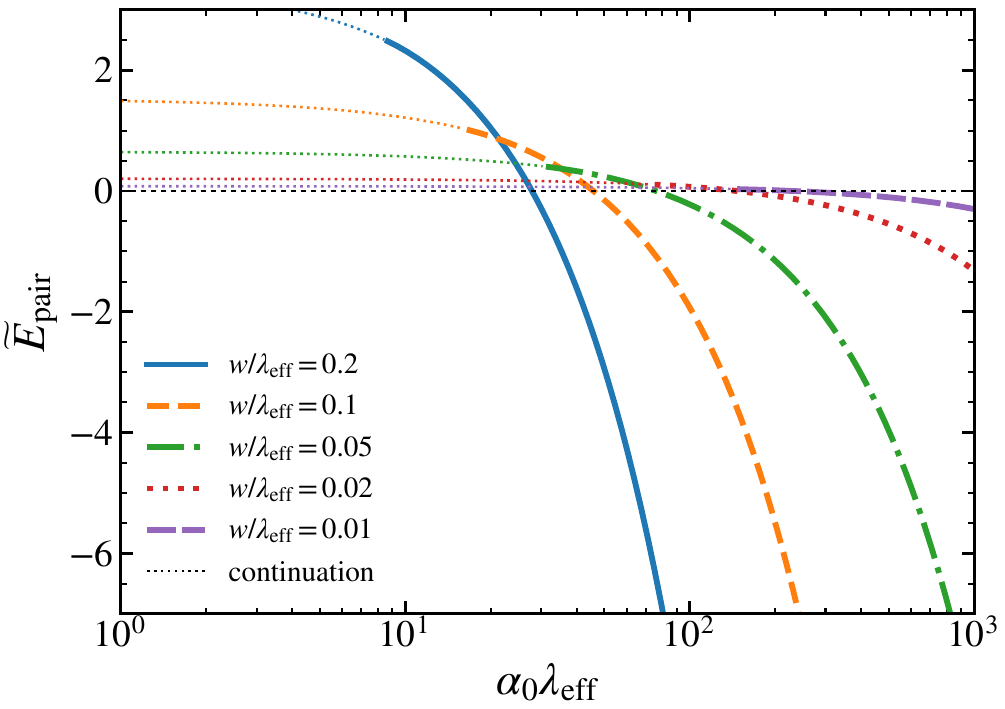}}}
\caption{Dependence of the dimensionless vortex--antivortex pair energy
$\widetilde{E}_{\mathrm{pair}}(x_\ast)$ on the
magnetoelectric coupling strength
$\alpha_0\lambda_{\mathrm{eff}}$ for several values of
$w/\lambda_{\mathrm{eff}}$. The energy
is evaluated at the equilibrium vortex position $x_{\ast }$ determined from
minimization of the full pair energy.}
\label{fig.08}
\end{figure}

Figure~\ref{fig.08} shows the dimensionless pair energy
$\widetilde{E}_{\mathrm{pair}}(u_*)$ evaluated at the equilibrium position.
The evolution with increasing magnetoelectric coupling separates naturally
into three regimes. At weak coupling, no finite-separation minimum exists
within the range where the logarithmic approximation is used. The dotted
curves in this regime are continuations obtained by evaluating
$E_{\mathrm{pair}}(u)$ at the smallest reference position retained within
the domain of validity of the calculation; they should therefore not be interpreted as equilibrium pair energies.

At intermediate coupling, a local minimum develops at finite separation,
\begin{equation}
\left.
\frac{dE_{\mathrm{pair}}}{du}
\right|_{u=u_*}
=
0,
\qquad
\left.
\frac{d^2E_{\mathrm{pair}}}{du^2}
\right|_{u=u_*}
>
0.
\end{equation}
The N\'{e}el wall can then support a bound vortex--antivortex configuration.
As long as
\begin{equation}
E_{\mathrm{pair}}(u_*)>0,
\end{equation}
however, the vortex-free state remains energetically favorable. The
finite-separation pair represents a metastable configuration rather than the
ground state.

Upon further increasing the magnetoelectric coupling, the minimum pair energy
reaches zero. We define the critical coupling $\alpha_c$ by
\begin{equation}
\min_u E_{\mathrm{pair}}(u;\alpha_c)
=
0.
\label{eq:critical-coupling}
\end{equation}
Equivalently,
\begin{equation}
E_{\mathrm{pair}}(u_*;\alpha_c)=0,
\qquad
\left.
\frac{\partial E_{\mathrm{pair}}}{\partial u}
\right|_{u=u_*,\,\alpha_0=\alpha_c}
=
0.
\label{eq:critical-coupling-system}
\end{equation}
For $\alpha_0>\alpha_c$, the minimum pair energy is negative relative to the
vortex-free state. Within the present London energy analysis, formation of a
vortex--antivortex pair is then energetically favorable even in zero applied
magnetic field.

Note that we consider temperatures below the possible
Berezinskii--Kosterlitz--Thouless (BKT) transition temperature,
$T_{\mathrm{BKT}}$, in thin-film superconductors. For films with
sufficiently large superfluid stiffness, $T_{\mathrm{BKT}}$ is usually
close to the mean-field transition temperature $T_{c0}$, and the relative
difference $(T_{c0}-T_{\mathrm{BKT}})/T_{c0}$ can be very small
(less than $1\%$ for films in the nanometer-thickness range)~\cite{Beasley1979}.
Therefore, this condition is not overly restrictive.
We estimate that for temperatures sufficiently below $T_{\mathrm{BKT}}$
the relative shift of the critical magnetoelectric coupling is of the
order of
\begin{equation}
\frac{\delta\alpha_c}{\alpha_c}
\sim
\frac{T_{c0}-T_{\mathrm{BKT}}}{T_{c0}-T}
\ll 1.
\end{equation}
Thus, the corresponding shift of $\alpha_c$ is expected to be small.

\begin{figure}[tb]
\resizebox{0.98\columnwidth}{!}{\rotatebox{0}{%
\includegraphics{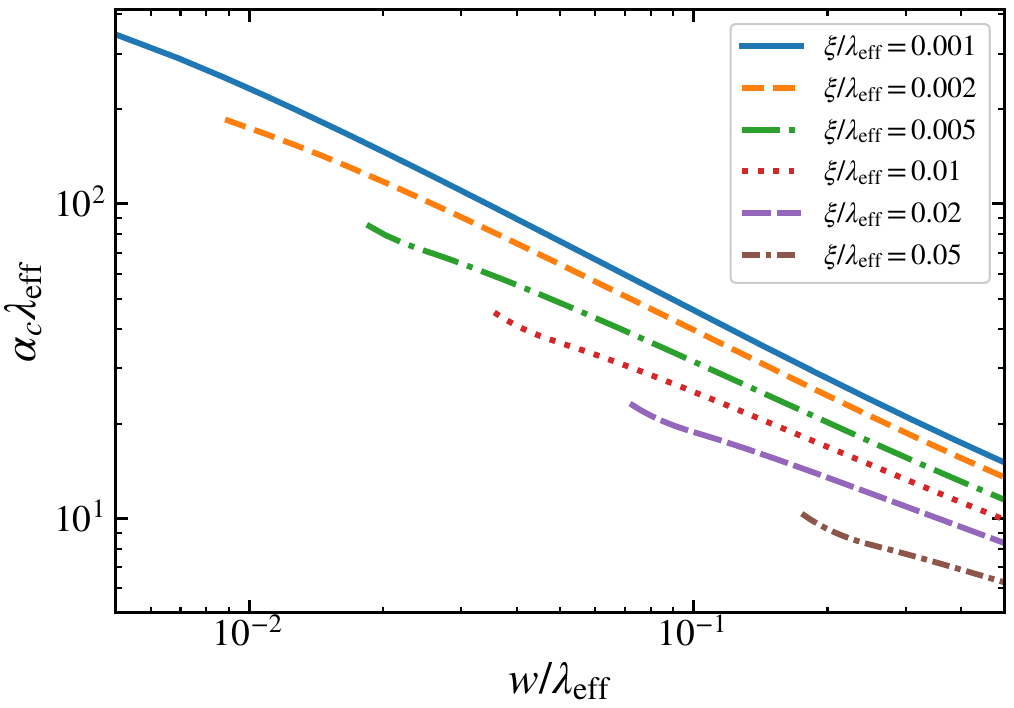}}}
\caption{Critical dimensionless magnetoelectric coupling $\protect\alpha _{c}%
\protect\lambda _{\mathrm{eff}}$ required for spontaneous vortex--antivortex
generation as a function of the ratio $w/\protect\lambda _{\mathrm{eff}}$
for several values of the coherence-length ratio $\protect\xi /\protect%
\lambda _{\mathrm{eff}}$. }
\label{fig.09}
\end{figure}

Increasing the wall width lowers the critical magnetoelectric coupling
because the extended wall texture produces a stronger vortex interaction
over a larger spatial region. Figure~\ref{fig.09} shows
$\alpha_c\lambda_{\mathrm{eff}}$ as a function of
$w/\lambda_{\mathrm{eff}}$ for several values of
$\xi/\lambda_{\mathrm{eff}}$. Below the critical line, the vortex-free state
has lower energy. Above it, the minimum-energy vortex--antivortex
configuration lies below the vortex-free state.

The dependence on $\xi/\lambda_{\mathrm{eff}}$ originates from the
logarithmic energy cost of the pair in
Eq.~(\ref{eq:Epair-no-wall}). A smaller coherence length increases
$\ln(2x/\xi)$ and therefore increases the magnetoelectric coupling required
to stabilize the pair. The critical line thus reflects the competition
between the logarithmic vortex--antivortex pair energy and the extended N\'{e}el-wall-induced pinning potential.

\begin{figure}[tb]
\centering
\includegraphics[width= 1.0\columnwidth]
{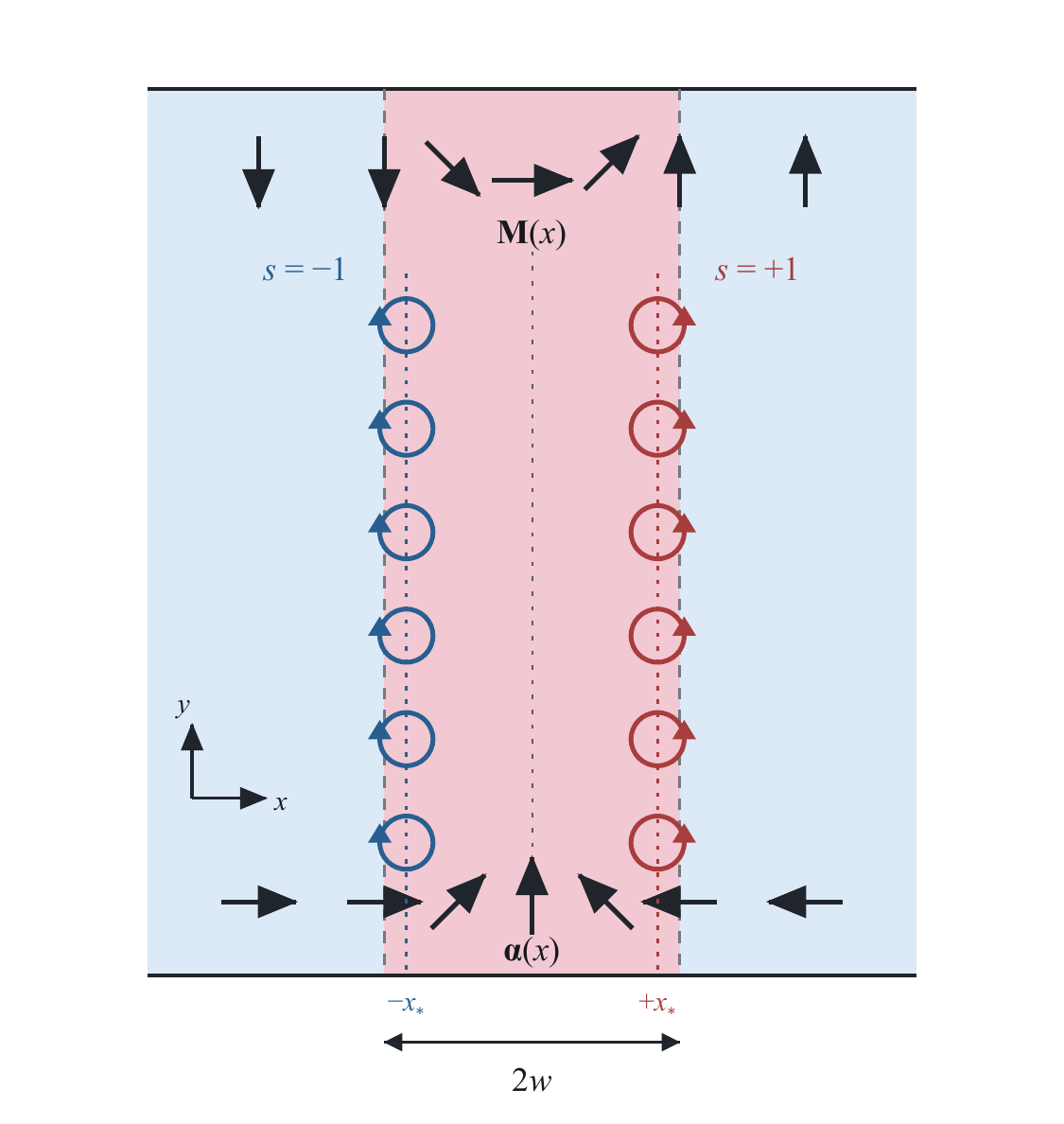}
\caption{Schematic decoration of a single N\'eel wall by parallel vortex and
antivortex chains in the dilute-pair limit. The arrows show the in-plane
profiles $\mathbf{M}(x)$ and $\boldsymbol{\alpha}(x)$. For
$\mathbf{n}=\hat{\mathbf{z}}$ and $\alpha_0>0$, vortices ($s=+1$) are
localized near $x=+x_*$ and antivortices ($s=-1$) near $x=-x_*$.}
\label{fig:neel-wall-multipair}
\end{figure}

\subsection{Parameter estimates and experimental feasibility}
We now compare the critical coupling obtained from the
vortex--antivortex calculation with the microscopic
estimate of the magnetoelectric coupling.

From Eq.~(8), we use
$|\alpha_0|\xi\sim10^{-2}$. The dimensionless coupling
that controls the vortex--wall interaction is therefore
$|\alpha_0|\lambda_{\rm eff}
=|\alpha_0|\xi/(\xi/\lambda_{\rm eff})$.
For representative thin-film values
$\xi/\lambda_{\rm eff}\sim10^{-3}-10^{-2}$, this gives
\begin{equation}
|\alpha_0|\lambda_{\rm eff}\sim1-10.
\end{equation}
More specifically,
$|\alpha_0|\lambda_{\rm eff}\sim10$ for
$\xi/\lambda_{\rm eff}=10^{-3}$, whereas
$|\alpha_0|\lambda_{\rm eff}\sim1$ for
$\xi/\lambda_{\rm eff}=10^{-2}$.

Figures~5 and 6 were calculated for
$\xi/\lambda_{\rm eff}=10^{-3}$. They show that, as the
magnetoelectric coupling increases, a finite-separation
minimum of the vortex--antivortex energy first appears
and then moves below the energy of the vortex-free state.
The required coupling becomes smaller for wider N\'eel
walls because the wall-induced pinning potential becomes
stronger and extends over a larger spatial region.

Figure~7 shows the corresponding critical coupling
$\alpha_c\lambda_{\rm eff}$. It decreases both with
increasing $w/\lambda_{\rm eff}$ and with increasing
$\xi/\lambda_{\rm eff}$. The latter effect follows from
the logarithmic vortex--antivortex energy: a larger
coherence length reduces the energy cost of forming the
pair. For extended N\'eel walls with
$w/\lambda_{\rm eff}\sim0.3-0.5$, Fig.~7 gives the
representative threshold
\begin{equation}
\alpha_c\lambda_{\rm eff}\sim5-15,
\end{equation}
depending on the value of $\xi/\lambda_{\rm eff}$.

It is important to compare this critical value with the
coupling expected from the microscopic estimate.
For $\xi/\lambda_{\rm eff}=10^{-3}$, one has
$|\alpha_0|\lambda_{\rm eff}\sim10$, which is of the same
order as the calculated threshold. By contrast, when
$\xi/\lambda_{\rm eff}$ increases to $10^{-2}$,
the estimated value of $|\alpha_0|\lambda_{\rm eff}$
decreases to approximately $1$. The corresponding
decrease of $\alpha_c\lambda_{\rm eff}$ in Fig.~7 is much
weaker. Thus, although the critical threshold itself
becomes lower when $\xi/\lambda_{\rm eff}$ increases,
the microscopic estimate of the magnetoelectric coupling
decreases even more strongly. Therefore, for fixed
$|\alpha_0|\xi\sim10^{-2}$, the regime
$\xi/\lambda_{\rm eff}\sim10^{-3}$ is more favorable for
reaching the vortex--antivortex nucleation threshold.

The required wall width can also be expressed in units
of the coherence length. Since
$w/\xi=(w/\lambda_{\rm eff})/
(\xi/\lambda_{\rm eff})$, the interval
$w/\lambda_{\rm eff}\sim0.3-0.5$ corresponds to
$w/\xi\sim300-500$ for
$\xi/\lambda_{\rm eff}=10^{-3}$ and to
$w/\xi\sim30-50$ for
$\xi/\lambda_{\rm eff}=10^{-2}$.
Thus, in both cases the relevant N\'eel walls are broad
compared with the superconducting coherence length,
$\xi\ll w$.

The wall width and the magnetoelectric coupling are
controlled by different microscopic mechanisms.
The wall width is determined by the magnetic exchange
stiffness, anisotropy, and magnetostatic energy, whereas
$\alpha_0$ is determined by the interfacial spin--orbit
and exchange coupling. In the present model, $w$ and
$\alpha_0$ are therefore treated as separate input
parameters. Increasing $\alpha_0$ does not imply a
corresponding increase of the wall width. Consequently,
the required values of $w$, $\lambda_{\rm eff}/\xi$, and
$\alpha_0$ must be realized simultaneously in the same
heterostructure.

YIG-based hybrids illustrate that several of the required
ingredients can be realized experimentally. N\'eel-type
domain walls with micrometer-scale widths have been
reported in YIG/Pt structures \cite{Mendil2019}, while superconducting
YIG/NbN hybrids and interfacial exchange--Rashba effects
in Pt/YIG-based structures have also been demonstrated \cite{Rogdakis2019}.
These results make YIG-based hybrids interesting candidate
platforms \cite{Jeon2022}. However, the cited experiments demonstrate
these ingredients separately and do not establish that
the required values of $w/\lambda_{\rm eff}$,
$\xi/\lambda_{\rm eff}$, and
$|\alpha_0|\lambda_{\rm eff}$ can all be realized in the
same heterostructure.

The most favorable experimental regime therefore
combines a very thin superconducting film with a large
$\lambda_{\rm eff}/\xi$, a broad N\'eel wall with
$w/\lambda_{\rm eff}$ of order a few tenths, and strong
interfacial magnetoelectric coupling. For the nominal
estimate $|\alpha_0|\xi\sim10^{-2}$,
$\xi/\lambda_{\rm eff}\sim10^{-3}$ gives
$|\alpha_0|\lambda_{\rm eff}\sim10$, which is comparable
to the calculated threshold
$\alpha_c\lambda_{\rm eff}\sim5-15$.
This identifies a target regime in which spontaneous
vortex--antivortex generation may become experimentally
accessible. Stronger Rashba spin--orbit coupling, larger
exchange fields, or a larger effective Pearl length would
further increase $|\alpha_0|\lambda_{\rm eff}$ and make
the nucleation condition easier to satisfy.

For $\alpha_0>\alpha_c$, a qualitative extension of the
single-pair result is possible because the wall-induced
potential is translationally invariant along $y$. In the
dilute-pair limit, several well-separated pairs may occupy
different longitudinal positions. The N\'eel wall is then
expected to be decorated by a vortex chain near
$x\simeq x^\ast$ and a parallel antivortex chain near
$x\simeq -x^\ast$, as illustrated schematically in
Fig.~\ref{fig:neel-wall-multipair}.

In our analysis, we treat the magnetization $M(x)$ as
fixed and neglect the feedback of superconductivity on
the magnetic structure. While the superconducting
condensation energy is usually significantly smaller than
the magnetic energy, backreaction effects may become
relevant in very thin magnetic films and alter the
underlying magnetic structure
(see, for instance, Refs.~\cite{Buzdin1988,Mironov2026}).
This mechanism deserves further investigation.

\section{Conclusions}
\label{sec:conclusions}

We have developed an analytical description of the superconducting response to 
a N\'{e}el domain wall in a Rashba-coupled superconductor--ferromagnet bilayer. 
We also considered a periodic cycloid as an independent realization of a continuously 
rotating planar magnetic texture. For the periodic cycloid, the linearized Ginzburg--Landau 
analysis yields a spatially modulated order-parameter profile at the onset of superconductivity. 
Below $T_c$, the corresponding constant-modulus London description yields a transverse 
magnetoelectric current, a sign-changing magnetic induction, and a vorticity-selective vortex force.

For the single N\'{e}el wall, the equilibrium phase gradient compensates the
longitudinal component of the magnetoelectric source, whereas the transverse
component remains uncompensated and generates a spontaneous sheet current
parallel to the wall. Pearl screening redistributes this current and produces
an out-of-plane magnetic induction whose symmetry is fixed by the wall
geometry: the screened current is even with respect to the wall center,
whereas the magnetic field is odd and changes sign across the wall.

The screened current generates an antisymmetric and vorticity-selective
vortex--wall interaction. For a given vorticity, one side of the wall is
attractive and the opposite side is repulsive, with the roles reversed for
an antivortex. Consequently, in zero applied magnetic field, the relevant
neutral excitation is a vortex--antivortex pair pinned on opposite attractive
branches of the wall potential. The central physical result is that the wall
does not merely pin pre-existing vortices: when the spin--orbit-induced
magnetoelectric coupling exceeds a critical value, the minimum energy of a
bound vortex--antivortex pair falls below that of the vortex-free state.

The threshold is controlled by
$|\alpha_0|\lambda_{\mathrm{eff}}$,
$w/\lambda_{\mathrm{eff}}$, and
$\xi/\lambda_{\mathrm{eff}}$. For the representative estimates
$|\alpha_0|\xi\sim10^{-2}$ and
$\xi/\lambda_{\mathrm{eff}}\sim10^{-3}$, one obtains
$|\alpha_0|\lambda_{\mathrm{eff}}\sim10$, comparable to the calculated
threshold
$\alpha_c\lambda_{\mathrm{eff}}\sim5\text{--}15$
for walls with
$w/\lambda_{\mathrm{eff}}\sim0.3\text{--}0.5$.
The most favorable regime is therefore expected in thin superconducting
films with a large ratio $\lambda_{\mathrm{eff}}/\xi$, strong interfacial
SOC, and wall widths satisfying
$\xi\ll w\lesssim\lambda_{\mathrm{eff}}$.

Although the explicit profiles were obtained in the SOC-dominated regime,
the London/Pearl formulation also allows the direct magnetostatic
contribution of the magnetic texture to be included in the same transverse
source. The spin--orbit and orbital contributions may therefore reinforce
or partially compensate each other, depending on the magnetic-texture
orientation and the interface geometry. The one-pair threshold derived here
uses the logarithmic vortex--antivortex interaction and is quantitatively
controlled when
$\xi\ll2x_*\ll\lambda_{\mathrm{eff}}$.

The oriented barrier--well potential provides the static asymmetry required for a vortex-ratchet response. In an array of N\'{e}el walls with the same potential orientation, a suitable zero-mean ac transport current may drive a vortex across the lower barrier during one half-cycle while it remains pinned during the opposite half-cycle, resulting in a net directional displacement. The attractive well also provides a mechanism for vortex dragging: as a N\'{e}el wall moves, it translates the pinning minimum, so that a vortex trapped in this minimum can be transported together with the wall.

Together with the energetic stabilization of zero-field vortex--antivortex pairs, these effects identify distinct routes for magnetic-texture control of vortex matter in superconducting films.

\begin{acknowledgments}
This work was supported by ANR-22-CE30-00020-01 ``SUPERFAST,'' the LIGHT
S\&T Graduate Program (PIA3 Investment for the Future Program, Grant
No.~ANR-17-EURE-0027), and GPR LIGHT. M.D.C. gratefully acknowledges
financial support and hospitality from the University of Bordeaux. He also
acknowledges the Federal University of Pernambuco for institutional support
through the PROPESQ--UFPE program (Edital PROPESQ No.~05/2018, Recife,
Brazil).
\end{acknowledgments}

\clearpage

\bibliography{references}

\end{document}


\title{Supplemental Material for ``Magnetoelectric response to a N\'eel
domain wall in spin--orbit-coupled superconducting films: spontaneous
currents, vortex pinning, and vortex--antivortex nucleation''}

\author{P. J. F. Cavalcanti}
\email{paulo-jose.fonseca-cavalcanti@u-bordeaux.fr}
\affiliation{University Bordeaux, LOMA UMR-CNRS 5798,
F-33405 Talence Cedex, France}

\author{M. D. Croitoru}
\email{mihail.croitoru@ufpe.br}
\affiliation{University Bordeaux, LOMA UMR-CNRS 5798,
F-33405 Talence Cedex, France}
\affiliation{Departamento de F\'{\i}sica, Centro de Ci\^encias Exatas e da
Natureza, Universidade Federal de Pernambuco, Recife, PE 50740-560, Brazil}

\author{A. I. Buzdin}
\email{alexandre.bouzdine@u-bordeaux.fr}
\affiliation{University Bordeaux, LOMA UMR-CNRS 5798,
F-33405 Talence Cedex, France}
\affiliation{Institute for Computer Science and Mathematical Modeling,
Sechenov First Moscow State Medical University, Moscow, Russia}

\date{\today}

\maketitle

\appendix


This Supplemental Material provides technical derivations
supporting the analytical results presented in the main text.
Appendix~\ref{app:harmonic-belowTc} derives the Fourier-space
Pearl screening kernel, screened London current, and vortex
interaction for the periodic harmonic cycloid.
Appendix~\ref{app:smooth-wall-check} examines the robustness
of the long-distance tails with respect to the wall profile.
Appendix~\ref{app:single-wall-pair} derives the single-wall
contribution to the vortex--antivortex pair energy,
the wall-edge estimate, and the equilibrium pair position.

\section{Periodic harmonic cycloid in the London/Pearl regime}
\label{app:harmonic-belowTc}

In this Appendix we derive the London response generated by a periodic harmonic
cycloid in the fixed-amplitude regime. We obtain the corresponding Pearl
screening kernel and the screened current distribution, which are used in the
main text to describe the interaction between the magnetoelectric current and a
vortex. Throughout this Appendix we consider temperatures sufficiently below
$T_c$, so that spatial variations of the condensate amplitude may be neglected
everywhere except inside vortex cores. The superconducting order parameter is
therefore written as
\begin{equation}
\psi(\mathbf r,T)
=
\Psi(T)e^{i\varphi(\mathbf r)},
\label{eq:london-order-parameter}
\end{equation}
where $\Psi(T)$ is the spatially uniform amplitude and
$\varphi(\mathbf r)$ is the superconducting phase.
The Ginzburg--Landau equation used in the main text is
\begin{equation}
a\psi+b|\psi|^2\psi+\gamma\mathbf{D}^2\psi
+\gamma\left[
2\boldsymbol{\alpha}\cdot\mathbf{D}\psi
-i(\boldsymbol{\nabla}\cdot\boldsymbol{\alpha})\psi
\right]
=
0,
\label{eq:GL-london-supp}
\end{equation}
where
$\mathbf{D}=-i\boldsymbol{\nabla}-(2e/\hbar c)\mathbf{A}$.
For the periodic harmonic cycloid,
$\alpha_x(x)=-\alpha_0\sin(px)$. 
Within the London approximation, the condensate amplitude is 
spatially uniform, and only the superconducting phase remains to be determined. 
The phase equation follows from the imaginary part of the Ginzburg--Landau equation. 
In the present one-dimensional geometry, the screening vector potential is purely 
transverse and therefore does not contribute to the longitudinal phase equation. 
Substituting Eq.~\eqref{eq:london-order-parameter} into Eq.~\eqref{eq:GL-london-supp} 
and taking the imaginary part, we obtain
\begin{equation}
\frac{d}{dx}
\left[
\frac{d\varphi}{dx}
-
\alpha_0\sin(px)
\right]
=
0.
\label{eq:phase-equation}
\end{equation}
Equation~\eqref{eq:phase-equation} is immediately integrated. 
In the absence of an externally imposed longitudinal transport current, 
the integration constant vanishes because the equilibrium longitudinal 
supercurrent must be zero. Using $p=\pi/(2w)$, we obtain
\begin{equation}
\varphi(x)
=
-\frac{\alpha_0}{p}\cos(px)
=
-\frac{2\alpha_0w}{\pi}
\cos\!\left(
\frac{\pi x}{2w}
\right),
\label{eq:london-phase}
\end{equation}
up to an irrelevant additive constant.

The equilibrium phase completely compensates the longitudinal magnetoelectric 
contribution. Consequently, the remaining current source is purely transverse. 
The sheet current therefore takes the form
\begin{equation}
\mathbf j_s
=
\frac{c}{4\pi\lambda_{\mathrm{eff}}}
\left[
\frac{\phi_0}{2\pi}
\left(
\boldsymbol{\nabla}\varphi
+
\boldsymbol{\alpha}
\right)
-
\mathbf A(\mathbf r,0)
\right],
\label{eq:sheet-current}
\end{equation}
where $\lambda_{\mathrm{eff}}=\lambda_L^2/d_s$. 
Using the phase solution (\ref{eq:london-phase}), 
the longitudinal contribution vanishes identically, 
leaving only the transverse (bare) driving term
\begin{equation}
S_y(x)
=
A_0\cos(px),
\qquad
A_0
=
\frac{\phi_0\alpha_0}{2\pi}.
\label{eq:bare-source}
\end{equation}

For completeness, we also include the magnetostatic field 
generated by a thin ferromagnetic layer of thickness $d_f$. 
On the superconducting side ($z<0$), the corresponding vector 
potential and magnetic field are
\begin{equation}
A_y^{\rm M}(x,z)
=
2\pi m d_f\cos(px)e^{pz},
\qquad
z<0,
\label{eq:magnetostatic-vector-potential}
\end{equation}
and
\begin{equation}
\begin{aligned}
\mathbf h^{\rm M}(x,z)
={}&
-2\pi p m d_f e^{pz}
\\
&\times
\left[
\cos(px)\hat{\mathbf x}
+
\sin(px)\hat{\mathbf z}
\right],
\qquad
z<0.
\end{aligned}
\label{eq:magnetostatic-field}
\end{equation}
Thus, the signed transverse source entering the London current is obtained
by the replacement
\begin{equation}
A_0
\longrightarrow
A_0-2\pi m d_f.
\label{eq:combined-source-replacement}
\end{equation}
The spin--orbit and magnetostatic contributions may therefore reinforce or
partially compensate each other, depending on the sign of $\alpha_0$ and on
the orientation of the magnetic texture relative to the interface normal.
The spin--orbit-dominated regime considered for the vortex--wall interaction
corresponds to
\begin{equation}
\frac{|\alpha_0|\phi_0}{2\pi}
\gg
2\pi m d_f.
\label{eq:SOC-dominated-condition}
\end{equation}
In the remainder of this Appendix, we focus on the SOC-dominated 
regime and therefore retain only the magnetoelectric spin--orbit contribution. 
This approximation allows the screening problem to be solved analytically 
while preserving the essential physics of the vortex--wall interaction.

We now derive the Pearl screening kernel explicitly. Because 
the London-Maxwell equations are linear, it is convenient to solve the screening 
problem in Fourier space, where each spatial Fourier component can be treated 
independently. This procedure leads directly to the characteristic Pearl 
screening factor governing thin superconducting films. In Coulomb gauge, the
transverse vector potential satisfies
\begin{equation}
-\left(
\partial_x^2+\partial_z^2
\right)
A_y(x,z)
=
\frac{1}{\lambda_{\mathrm{eff}}}
\left[
S_y(x)-A_y(x,0)
\right]\delta(z).
\label{eq:pearl-maxwell-supp}
\end{equation}
Introducing the Fourier representations
\begin{equation}
\begin{aligned}
A_y(x,z)
&=
\int\frac{dq}{2\pi}
\int\frac{dk}{2\pi}
A(q,k)e^{i(qx+kz)},
\\
A_q
&=
\int\frac{dk}{2\pi}A(q,k),
\\
S_q
&=
\int dx\,e^{-iqx}S_y(x).
\end{aligned}
\label{eq:pearl-fourier-definitions-supp}
\end{equation}
Equation~\eqref{eq:pearl-maxwell-supp} then becomes
\begin{equation}
\left(q^2+k^2\right)A(q,k)
=
\frac{1}{\lambda_{\mathrm{eff}}}
\left(
S_q-A_q
\right).
\label{eq:pearl-fourier-equation-supp}
\end{equation}
Integrating over $k$ and using
$\int dk/[2\pi(q^2+k^2)]=1/(2|q|)$ gives
\begin{equation}
A_q
=
\frac{S_q}
{1+2|q|\lambda_{\mathrm{eff}}}.
\label{eq:pearl-kernel-supp}
\end{equation}
Equation~\eqref{eq:pearl-kernel-supp} is the standard Pearl 
screening kernel for a thin superconducting film. It shows that each Fourier component 
of the driving source is reduced by the factor $(1+2|q|\lambda_{\mathrm{eff}})^{-1}$, 
reflecting the nonlocal magnetic screening characteristic of Pearl electrodynamics.

For the single harmonic $S_y(x)=A_0\cos(px)$ with $p>0$,
Eq.~\eqref{eq:pearl-kernel-supp} gives
\begin{equation}
A_y(x,z)
=
\frac{A_0}
{1+2p\lambda_{\mathrm{eff}}}
\cos(px)e^{-p|z|}.
\label{eq:screened-vector-potential}
\end{equation}
Consequently, the screened sheet current is
\begin{equation}
j_y(x)
=
j_0
\frac{2p\lambda_{\mathrm{eff}}}
{1+2p\lambda_{\mathrm{eff}}}
\cos(px),
\qquad
j_0
=
\frac{cA_0}{4\pi\lambda_{\mathrm{eff}}},
\label{eq:screened-current}
\end{equation}
The screening modifies only the amplitude of the harmonic current distribution, 
while its spatial dependence remains unchanged. In the limit $p\lambda_{\mathrm{eff}}\ll1$, 
magnetic screening strongly suppresses the current, whereas for $p\lambda_{\mathrm{eff}}\gg1$ 
the current approaches the unscreened London result. The screening field generated by 
the superconducting sheet current is
\begin{equation}
h_z^{\rm sc}(x,0)
=
-\frac{pA_0}
{1+2p\lambda_{\mathrm{eff}}}
\sin(px).
\label{eq:screening-field}
\end{equation}

As an independent derivation of the vortex interaction, 
we separate the superconducting phase into a regular contribution associated 
with the cycloidal magnetoelectric texture and a singular contribution 
generated by the vortex. Correspondingly, we introduce the regular 
and vortex source fields
\begin{equation}
\begin{aligned}
\mathbf C(\mathbf r)
&=
\frac{\phi_0}{2\pi}
\left[
\boldsymbol{\nabla}\varphi_{\mathrm{reg}}(\mathbf r)
+
\boldsymbol{\alpha}(\mathbf r)
\right],
\\
\boldsymbol{\Phi}(\mathbf r)
&=
\frac{\phi_0}{2\pi}
\boldsymbol{\nabla}\varphi_{\mathrm{vor}}(\mathbf r).
\end{aligned}
\label{eq:wall-vortex-sources-supp}
\end{equation}
Within the fixed-amplitude London approximation, 
the free energy 
can be written as
\begin{equation}
\begin{aligned}
E_{\rm L}
={}&
\frac{1}{8\pi}
\int d^3r\,\mathbf h^2
\\
&+
\frac{1}{8\pi\lambda_{\mathrm{eff}}}
\int d^2r\,
\left[
\mathbf C
+
\boldsymbol{\Phi}
-
\mathbf A(\mathbf r,0)
\right]^2,
\end{aligned}
\label{eq:london-energy-supp}
\end{equation}
up to terms independent of the phase and vector potential. 
 Using the Pearl kernel, the vector potential 
can be eliminated analytically. The interaction energy 
between the regular magnetoelectric source and 
the vortex source then becomes
\begin{equation}
\begin{aligned}
E_{\mathrm{int}}
={}&
\frac{1}{4\pi}
\int\frac{d^2q}{(2\pi)^2}
\frac{|\mathbf q|}
{1+2\lambda_{\mathrm{eff}}|\mathbf q|}
\\
&\times
\left[
\mathbf C_{\mathbf q}\cdot
\boldsymbol{\Phi}_{\mathbf q}^{*}
+
\mathbf C_{\mathbf q}^{*}\cdot
\boldsymbol{\Phi}_{\mathbf q}
\right].
\end{aligned}
\label{eq:interaction-kernel-supp}
\end{equation}

For the periodic harmonic cycloid and a single vortex of vorticity $s=\pm1$
centered at $\mathbf r_0=(x_0,y_0)$, the corresponding 
Fourier components take the form
\begin{equation}
\begin{aligned}
\mathbf C_{\mathbf q}
={}&
\phi_0\alpha_0\pi
\left[
\delta(q_x-p)+\delta(q_x+p)
\right]
\delta(q_y)\hat{\mathbf y},
\\
\boldsymbol{\Phi}_{\mathbf q}
={}&
is\phi_0
e^{-i\mathbf q\cdot\mathbf r_0}
\frac{
q_y\hat{\mathbf x}
-
q_x\hat{\mathbf y}
}
{|\mathbf q|^2}.
\end{aligned}
\label{eq:harmonic-vortex-sources-supp}
\end{equation}
Substitution into Eq.~\eqref{eq:interaction-kernel-supp} gives
\begin{equation}
E_{\rm int}^{(s)}(x_0)
=
-s\frac{\phi_0^2\alpha_0}{4\pi^2}
\frac{\sin(px_0)}
{1+2p\lambda_{\mathrm{eff}}}.
\label{eq:harmonic-vortex-interaction}
\end{equation}
The interaction energy is sinusoidal along the cycloid and 
changes sign with the vortex vorticity. Magnetic screening preserves 
the spatial dependence while reducing the interaction amplitude 
through the Pearl factor $(1+2p\lambda_{\mathrm{eff}})^{-1}$.
The result is independent of $y_0$, as required by translational invariance
along the $y$ direction.

As a consistency check, differentiating Eq.~\eqref{eq:harmonic-vortex-interaction}, 
with respect to the vortex position gives
\begin{equation}
-\frac{dE_{\rm int}^{(s)}}{dx_0}\hat{\mathbf x}
=
\frac{s\phi_0}{c}
j_y(x_0)\hat{\mathbf x},
\label{eq:Lorentz-check}
\end{equation}
which is the Lorentz force exerted by the screened sheet current on the
vortex.

\section{Robustness of the long-distance Pearl tail with respect to the wall profile}
\label{app:smooth-wall-check}
The single-wall calculation in the main text uses a piecewise N\'eel-wall
profile because it allows an analytical treatment. The magnetization is
continuous at $x=\pm w$, but its first derivative is discontinuous. It is
therefore important to check that the algebraic long-distance current found
in the main text is not caused by these derivative discontinuities.

For the piecewise profile, the transverse magnetoelectric source is
\begin{equation}
\alpha_{{\rm pw},y}(x)
=
\begin{cases}
\alpha_0\cos\!\left(\dfrac{\pi x}{2w}\right), & |x|\leq w,\\[6pt]
0, & |x|>w.
\end{cases}
\label{eq:alpha-pw-smooth-check}
\end{equation}
Its Fourier transform is
\begin{equation}
\alpha_{{\rm pw},y}(k_x)
=
\frac{\alpha_0\pi}{w}
\frac{\cos(k_xw)}{
\left(\dfrac{\pi}{2w}\right)^2-k_x^2}.
\label{eq:alpha-pw-fourier-smooth-check}
\end{equation}
For $|k_x|w\gg1$,
\begin{equation}
\alpha_{{\rm pw},y}(k_x)
\simeq
-\frac{\alpha_0\pi}{w}
\frac{\cos(k_xw)}{k_x^2}.
\label{eq:alpha-pw-large-k-smooth-check}
\end{equation}
Thus, the piecewise profile has an algebraic $k_x^{-2}$ tail at large
$|k_x|$. This behavior comes from the jumps of
$d\alpha_{{\rm pw},y}/dx$ at $x=\pm w$. 

We now repeat the calculation for the smooth micromagnetic profile
\begin{equation}
\mathbf M_{\rm sm}(x)
=
m
\left[
\operatorname{sech}\!\left(\frac{x}{2w}\right),
\tanh\!\left(\frac{x}{2w}\right),
0
\right].
\label{eq:M-smooth-check}
\end{equation}
Using the same convention
$\boldsymbol{\alpha}=\alpha_0(\mathbf n\times\mathbf M)/m$, the transverse
source is
\begin{equation}
\alpha_{{\rm sm},y}(x)
=
\alpha_0
\operatorname{sech}\!\left(\frac{x}{2w}\right).
\label{eq:alpha-smooth-y-check}
\end{equation}
Using
\begin{equation}
\int_{-\infty}^{\infty}ds\,
\operatorname{sech}(s)e^{-iqs}
=
\frac{\pi}{\cosh(\pi q/2)},
\label{eq:sech-transform-check}
\end{equation}
we obtain
\begin{equation}
\alpha_{{\rm sm},y}(k_x)
=
2\pi w\alpha_0
\operatorname{sech}(\pi wk_x).
\label{eq:alpha-smooth-fourier-check}
\end{equation}
For $|k_x|w\gg1$,
\begin{equation}
\alpha_{{\rm sm},y}(k_x)
\simeq
4\pi w\alpha_0 e^{-\pi w|k_x|}.
\label{eq:alpha-smooth-large-k-check}
\end{equation}
Thus, unlike the piecewise wall, the smooth wall has an exponentially
decaying high-wave-vector spectrum.

For both profiles, the Pearl-screened sheet current can be written as
\begin{equation}
j_y(k_x)
=
j_y^{(0)}(k_x)
\frac{2\lambda_{\mathrm{eff}}|k_x|}
{1+2\lambda_{\mathrm{eff}}|k_x|},
\label{eq:screened-current-general-check}
\end{equation}
where $j_y^{(0)}(k_x)$ is the Fourier transform of the unscreened current.
The long-distance behavior in real space is determined by the small-$|k_x|$
part of this expression, not by the large-$|k_x|$ behavior of the wall
profile.
For any localized even source with finite integral
\begin{equation}
I_0
=
\int_{-\infty}^{\infty}dx\,
\frac{j_y^{(0)}(x)}{j_0},
\label{eq:I0-general-check}
\end{equation}
we have at small $|k_x|$
\begin{equation}
j_y^{(0)}(k_x)
=
j_0 I_0+O(k_x^2),
\label{eq:source-small-k-check}
\end{equation}
and therefore
\begin{equation}
j_y(k_x)
=
2j_0 I_0\lambda_{\mathrm{eff}}|k_x|
+O(k_x^2).
\label{eq:current-small-k-check}
\end{equation}
For $x\neq0$,
\begin{equation}
\int_{-\infty}^{\infty}
\frac{dk_x}{2\pi}\,
|k_x|e^{ik_xx}
=
-\frac{1}{\pi x^2},
\label{eq:abs-k-transform-check}
\end{equation}
so that
\begin{equation}
j_y(x)
\simeq
-\frac{2j_0 I_0\lambda_{\mathrm{eff}}}{\pi x^2},
\qquad
|x|\gg w,\lambda_{\mathrm{eff}}.
\label{eq:general-tail-check}
\end{equation}
Thus, the $x^{-2}$ far-field behavior follows from the small-$|k_x|$
nonanalyticity of the Pearl factor.

For the piecewise profile,
\begin{equation}
I_0^{\rm pw}
=
\int_{-w}^{w}dx\,
\cos\!\left(\frac{\pi x}{2w}\right)
=
\frac{4w}{\pi},
\label{eq:I0-pw-check}
\end{equation}
which gives
\begin{equation}
\frac{j_{{\rm pw},y}(x)}{j_0}
\simeq
-\frac{8}{\pi^2}
\frac{w\lambda_{\mathrm{eff}}}{x^2}.
\label{eq:tail-pw-check}
\end{equation}
This agrees with the leading asymptotic result used in the main text.

For the smooth profile,
\begin{equation}
I_0^{\rm sm}
=
\int_{-\infty}^{\infty}dx\,
\operatorname{sech}\!\left(\frac{x}{2w}\right)
=
2\pi w,
\label{eq:I0-sm-check}
\end{equation}
and therefore
\begin{equation}
\frac{j_{{\rm sm},y}(x)}{j_0}
\simeq
-4\frac{w\lambda_{\mathrm{eff}}}{x^2}.
\label{eq:tail-sm-check}
\end{equation}
The numerical coefficient is different because the integrated source is
different for the two profiles, but the $x^{-2}$ dependence is the same.

This comparison shows clearly that the two effects come from different
parts of Fourier space. The derivative discontinuities of the piecewise wall
control the large-$|k_x|$ behavior and therefore influence the current close
to the wall. In contrast, the long-distance current is controlled by the
small-$|k_x|$ Pearl factor. The algebraic far-field tail is therefore a
property of Pearl electrodynamics and remains present for the smooth wall,
even though its high-$|k_x|$ spectrum decays exponentially. The same
argument applies to the long-distance magnetic field and vortex-interaction
kernels obtained from the same screened source.

\section{Single N\'eel wall: vortex--antivortex pair energetics}
\label{app:single-wall-pair}

\subsection{Interaction energy at the wall edge}

This Appendix derives the interaction energy between a single N\'eel 
wall and a vortex located at the wall edge. The resulting expression is 
then used to determine the condition for spontaneous vortex--antivortex nucleation.
For a single N\'eel wall, the interaction energy between the wall-induced 
magnetoelectric texture and a vortex located at the wall edge, $x=w$, is
\begin{equation}
E_{\mathrm{int}}\left( w\right) =-\frac{\varepsilon m}{\gamma}\frac{\phi
_{0}^{2}}{(2\pi)^{3}}\,\mathcal{I}\left( w\right) ,
\end{equation}
\begingroup
\hbadness=10000
where $\mathcal{I}\left( w\right) =2\left[ J\left( 2w\right) +J\left(
0\right) \right] $. Since $J\left( s\right) =\mathrm{sgn}\left( s\right)
J\left( \left\vert s\right\vert \right) $, and directly from the integral
representation one has $J\left( 0\right) =0$, we obtain $\mathcal{I}\left(
w\right) =2J\left( 2w\right) $. Therefore,
\par
\endgroup

\begin{equation}
E_{\mathrm{int}}\left( w\right) =-\frac{\varepsilon m}{\gamma}\frac{\phi
_{0}^{2}}{(2\pi)^{3}}\,2J\left( 2w\right) .
\end{equation}
Using%
\begin{equation}
\begin{aligned}
J\left( s\right) & =\frac{\frac{\pi\lambda_{\mathrm{eff}}}{w}F\left( \frac{w%
}{\pi\lambda_{\mathrm{eff}}}\frac{\pi s}{2w}\right) -F\left( \frac{\pi s}{2w}%
\right) }{\left( \frac{\pi\lambda_{\mathrm{eff}}}{w}\right) ^{2}-1} \\
& -\frac{\pi\cos\left( \frac{\pi s}{2w}\right) }{2\left( \frac{\pi \lambda_{%
\mathrm{eff}}}{w}+1\right) },
\end{aligned}
\end{equation}
we obtain%
\begin{equation}
\begin{aligned}
J\left( 2w\right) & =\frac{\frac{\pi\lambda_{\mathrm{eff}}}{w}F\left( \frac{w%
}{\lambda_{\mathrm{eff}}}\right) -F\left( \pi\right) }{\left( \frac{%
\pi\lambda_{\mathrm{eff}}}{w}\right) ^{2}-1} \\
& +\frac{\pi}{2\left( \frac{\pi\lambda_{\mathrm{eff}}}{w}+1\right) },
\end{aligned}
\end{equation}
Thus, the interaction energy is given by
\begin{equation}
\begin{aligned}
E_{\mathrm{int}}\left( w\right) & =-\frac{\varepsilon m}{\gamma}\frac {%
\phi_{0}^{2}}{4\pi^{3}}\left[ \frac{\frac{\pi\lambda_{\mathrm{eff}}}{w}%
F\left( \frac{w}{\lambda_{\mathrm{eff}}}\right) -F\left( \pi\right) }{\left( 
\frac{\pi\lambda_{\mathrm{eff}}}{w}\right) ^{2}-1}\right.
\\
& +\left. \frac{\pi}{2\left( \frac{\pi\lambda_{\mathrm{eff}}}{w}+1\right) }%
\right] ,
\end{aligned}
\label{eq:Einta_final}
\end{equation}
Here

\begin{equation}
F\left( x\right) =\sin\left( x\right) \,\mathrm{Ci}\left( x\right)
-\cos\left( x\right) \,\mathrm{Si}\left( x\right) +\frac{\pi}{2}\cos\left(
x\right) ,
\end{equation}
and hence $F\left( \pi\right) =\mathrm{Si}\left( \pi\right) -\frac{\pi}{2}$.

To obtain the asymptotic form for $\lambda_{\mathrm{eff}}\gg w$, we introduce
\begin{equation}
\frac{1}{\Lambda}=\frac{w}{\lambda_{\mathrm{eff}}},\qquad\frac{1}{\Lambda}%
\ll1.
\end{equation}

For small argument,
\begin{equation}
F\left( \frac{1}{\Lambda}\right) =\frac{\pi}{2}+\frac{1}{\Lambda}\left[
\gamma_{E}-\ln\Lambda-1\right] +O\left( \Lambda^{-2}\right) ,
\end{equation}
where $\gamma_{E}$ is the Euler constant. Substituting these expansions into
the exact expression gives
\begin{equation}
\begin{aligned}
E_{\mathrm{int}}\left(\Lambda\right)\!\!
={}&\!-\frac{\varepsilon m}{\gamma}\!
\frac{\phi_{0}^{2}}{(2\pi)^{3}}\!
\left\{
\begin{gathered}
\frac{2}{\Lambda}
+\frac{1}{\Lambda^{2}}
\left[
\frac{2}{\pi}\left(\gamma_{E}-\ln\Lambda-1\right)
\right.\\
\left.
-\frac{2\,\mathrm{Si}\left(\pi\right)}{\pi^{2}}
\right]\\
+O\left[\Lambda^{-3}\ln\frac{1}{\Lambda}\right]
\end{gathered}
\right\}.
\end{aligned}
\end{equation}
Therefore, to leading order in $w/\lambda_{\mathrm{eff}}$,
\begin{equation}
E_{\mathrm{int}}\left( w\right) \simeq -\frac{\varepsilon m}{\gamma }\frac{%
\phi _{0}^{2}}{4\pi ^{3}}\frac{w}{\lambda _{\mathrm{eff}}}.
\label{main:S36cc}
\end{equation}%
The corresponding vortex--antivortex pair energy is  
\begin{equation}
E_{\mathrm{pair}}=\frac{\phi _{0}^{2}}{8\pi ^{2}\lambda _{\mathrm{eff}}}\ln
\left( \frac{2w}{\xi }\right) -\alpha _{0}\frac{\phi _{0}^{2}}{4\pi ^{3}}%
\frac{w}{\lambda _{\mathrm{eff}}}.
\end{equation}

\begingroup
\hbadness=10000
Spontaneous vortex--antivortex generation occurs when the pair energy
becomes negative: 
\begin{equation}
E_{\mathrm{pair}}<0.  \label{main:S37cc}
\end{equation}%
Using \cref{main:S36cc}, this condition gives 
\begin{equation}
\frac{\pi }{2\lambda _{\mathrm{eff}}}\ln \left( \frac{2w}{\xi }\right)
<\alpha _{0}\frac{w}{\lambda _{\mathrm{eff}}}.  \label{main:S40cc}
\end{equation}%
The threshold value of the magnetoelectric coupling is 
\begin{equation}
\alpha _{c}^{\mathrm{pair}}\left( w\right) =\frac{\pi }{2w}\ln \left( \frac{%
2w}{\xi }\right) .  \label{main:S42cc}
\end{equation}%
Spontaneous generation of a vortex--antivortex pair becomes energetically
favorable when 
\begin{equation}
\alpha_0 >\alpha _{c}^{\mathrm{pair}}\left( w\right) .  \label{main:S44cc}
\end{equation}
\par
\endgroup

\subsection{Optimal position of the vortex--antivortex pair}
\label{AppendixB}

The optimal position of the vortex--antivortex pair is obtained 
by minimizing the full pair energy. We consider a symmetric configuration 
with the vortex and antivortex located on opposite sides of the N\'eel wall,
\begin{equation}
\mathbf{R}_{\mathrm{vor}}=x\hat{\mathbf{x}},\qquad\mathbf{R}_{\mathrm{av}}=-x%
\hat{\mathbf{x}},  \label{app:pair_pos_01}
\end{equation}
where $x>0$. The separation $R$ should be understood as the distance between
the vortex and antivortex positions selected by the wall-induced pinning
potential. For a symmetric configuration with the vortex and antivortex
located at opposite sides of the wall, one may write 
\begin{equation}
R=\left\vert \mathbf{R}_{\mathrm{vor}}-\mathbf{R}_{\mathrm{av}}\right\vert
=2x.  \label{app:pair_pos_02}
\end{equation}
More generally, the equilibrium separation is determined by minimizing the full pair energy 
\begin{equation}
E_{\mathrm{pair}}\left( R\right) =E_{\mathrm{vor}}+E_{\mathrm{av}}+E_{%
\mathrm{int}}^{\mathrm{vor}}\left( R_{\mathrm{vor}}\right) +E_{\mathrm{int}%
}^{\mathrm{av}}\left( R_{\mathrm{av}}\right) +U_{\mathrm{v\bar{v}}}\left(
R\right)
\end{equation}
with respect to the vortex and antivortex positions. Here $R_{\mathrm{vor}%
}=x $, $R_{\mathrm{av}}=-x$ and $R=\left\vert \mathbf{R}_{\mathrm{vor}}-%
\mathbf{R}_{\mathrm{av}}\right\vert =2x$

Thus, the full pair energy is 
\begin{equation}
\begin{aligned}
E_{\mathrm{pair}}\left( x\right) & =E_{\mathrm{vor}}+E_{\mathrm{av}}
\\
& +E_{\mathrm{int}}^{\mathrm{vor}}\left( x\right) +E_{\mathrm{int}}^{\mathrm{%
av}}\left( -x\right) +U_{\mathrm{v\bar{v}}}\left( 2x\right) .
\end{aligned}
\label{app:pair_pos_03}
\end{equation}
The self-energies $E_{\mathrm{vor}}$ and $E_{\mathrm{av}}$ do not depend on
the position $x$, and therefore they do not enter the minimization condition.

The vortex interaction energy with the magnetoelectric texture is 
\begin{equation}
E_{\mathrm{int}}^{\mathrm{vor}}\left( x\right) =-{\alpha_0}\frac{\phi_{0}^{2}}{%
\left( 2\pi\right) ^{3}}\mathcal{I}\left( x\right) ,  \label{app:pair_pos_04}
\end{equation}
where ${\alpha_0}=\varepsilon m/\gamma$, and 
\begin{equation}
\mathcal{I}\left( x\right) =2\left[ J\left( x+w\right) +J\left( x-w\right) %
\right] .  \label{app:pair_pos_06}
\end{equation}
The kernel $J\left( s\right) $ is odd, $J\left( -s\right) =-J\left( s\right) 
$, and hence 
\begin{equation}
\mathcal{I}\left( -x\right) =-\mathcal{I}\left( x\right) .
\label{app:pair_pos_08}
\end{equation}

Because the antivortex has opposite vorticity, 
its phase-gradient source changes sign relative to that of the vortex. 
Since the magnetoelectric interaction is linear in the vortex source, the antivortex
interaction energy is $E_{\mathrm{int}}^{\mathrm{av}}\left( X\right) =-E_{%
\mathrm{int}}^{\mathrm{vor}}(X)$. For the antivortex placed at $X=-x$, we
obtain $E_{\mathrm{int}}^{\mathrm{av}}\left( -x\right) =-E_{\mathrm{int}}^{%
\mathrm{vor}}\left( -x\right) $. Using \cref{app:pair_pos_04,app:pair_pos_08}, $E_{\mathrm{int}}^{\mathrm{vor}}\left( -x\right) =-E_{%
\mathrm{int}}^{\mathrm{vor}}\left( x\right) $. Therefore 
\begin{equation}
E_{\mathrm{int}}^{\mathrm{av}}\left( -x\right) =E_{\mathrm{int}}^{\mathrm{vor%
}}\left( x\right) .  \label{app:pair_pos_12}
\end{equation}
Thus, for the symmetric vortex--antivortex configuration, both objects gain
the same magnetoelectric binding energy. Hence 
\begin{equation}
E_{\mathrm{int}}^{\mathrm{vor}}\left( x\right) +E_{\mathrm{int}}^{\mathrm{av}%
}\left( -x\right) =2E_{\mathrm{int}}^{\mathrm{vor}}\left( x\right) .
\label{app:pair_pos_13}
\end{equation}

The full pair energy can therefore be written as 
\begin{equation}
E_{\mathrm{pair}}\left( x\right) =E_{\mathrm{vor}}+E_{\mathrm{av}}+2E_{%
\mathrm{int}}^{\mathrm{vor}}\left( x\right) +U_{\mathrm{v\bar{v}}}\left(
2x\right) .  \label{app:pair_pos_14}
\end{equation}
Substituting \cref{app:pair_pos_04}, we obtain 
\begin{equation}
\begin{aligned}
E_{\mathrm{pair}}\left( x\right) & =E_{\mathrm{vor}}+E_{\mathrm{av}}
\\
& -2{\alpha_0}\frac{\phi_{0}^{2}}{\left( 2\pi\right) ^{3}}\mathcal{I}\left(
x\right) +U_{\mathrm{v\bar{v}}}\left( 2x\right) .
\end{aligned}
\label{app:pair_pos_15}
\end{equation}
where 
\begin{equation}
E_{\mathrm{vor}}=E_{\mathrm{av}}=\frac{\phi_{0}^{2}}{16\pi^{2}\lambda _{%
\mathrm{eff}}}\ln\left( \frac{\lambda_{\mathrm{eff}}}{\xi}\right) ,
\end{equation}
where $\lambda_{\mathrm{eff}}$ is the effective two-dimensional screening
length and $\xi$ is the coherence length. The vortex--antivortex interaction
is attractive. For separations $\xi\ll2x\ll\lambda_{\mathrm{eff}}$, its
logarithmic form is 
\begin{equation}
U_{\mathrm{v\bar{v}}}\left( 2x\right) =-\frac{\phi_{0}^{2}}{8\pi^{2}\lambda_{%
\mathrm{eff}}}\ln\left( \frac{\lambda_{\mathrm{eff}}}{2x}\right) .
\end{equation}

The optimal position is obtained from $dE_{\mathrm{pair}}/dx=0$. Since the
self-energies are constants, differentiation of \cref{app:pair_pos_15}
gives 
\begin{equation}
-2{\alpha_0}\frac{\phi_{0}^{2}}{\left( 2\pi\right) ^{3}}\frac{d\mathcal{I}\left(
x\right) }{dx}+\frac{d}{dx}U_{\mathrm{v\bar{v}}}\left( 2x\right) =0.
\label{app:pair_pos_17}
\end{equation}
The equilibrium condition becomes 
\begin{equation}
{\alpha_0}\frac{\phi_{0}^{2}}{\left( 2\pi\right) ^{3}}\frac{d\mathcal{I}\left(
x\right) }{dx}=\left. \frac{dU_{\mathrm{v\bar{v}}}\left( R\right)}{dR}%
\right\vert _{R=2x}.  \label{app:pair_pos_19}
\end{equation}
The right-hand side is positive for an attractive vortex--antivortex
interaction, because the energy increases when the two objects are pulled
apart.

Using \cref{app:pair_pos_06}, and taking into account the fact that for 
$-w<x<w$, one has $x+w>0$ and $x-w<0$. Since $J\left( s\right) $ is odd, $%
J^{\prime}\left( s\right) $ is even. Therefore $J^{\prime}\left( x-w\right)
=J^{\prime}\left( w-x\right) $, and 
\begin{equation}
\frac{d\mathcal{I}\left( x\right) }{dx}=2\left[ J^{\prime}\left( w+x\right)
+J^{\prime}\left( w-x\right) \right] .  \label{app:pair_pos_22}
\end{equation}
The exact implicit equation for the optimal pair position is therefore 
\begin{equation}
2{\alpha_0}\frac{\phi_{0}^{2}}{\left( 2\pi\right) ^{3}}\left[ J^{\prime}\left(
w+x\right) +J^{\prime}\left( w-x\right) \right] =\left. \frac {dU_{\mathrm{v%
\bar{v}}}\left( R\right) }{dR}\right\vert _{R=2x}.  \label{app:pair_pos_23}
\end{equation}
This is the general equation determining the optimal symmetric position of
the vortex--antivortex pair, $x=x_{\ast}$.

For $s>0$, the kernel takes the form 
\begin{equation}
\begin{aligned}
J\left( s\right) & =\frac{\frac{\pi\lambda_{\mathrm{eff}}}{w}F\left( \frac{w%
}{\pi\lambda_{\mathrm{eff}}}\frac{\pi s}{2w}\right) -F\left( \frac{\pi s}{2w}%
\right) }{\left( \frac{\pi\lambda_{\mathrm{eff}}}{w}\right) ^{2}-1}
\\
& -\frac{\pi\cos\left( \frac{\pi s}{2w}\right) }{2\left( \frac{\pi \lambda_{%
\mathrm{eff}}}{w}+1\right) }
\end{aligned}
\label{app:pair_pos_24}
\end{equation}
where%
\begin{equation}
F\left( u\right) =\sin u\,\mathrm{Ci}\left( u\right) -\cos u\left[ \mathrm{Si%
}\left( u\right) -\frac{\pi}{2}\right] .  \label{app:pair_pos_26}
\end{equation}
Then for $s>0$ 
\begin{equation}
\begin{aligned}
\frac{J^{\prime}\left( s\right) }{\frac{\pi}{2w}} & =\frac{G\left( \frac{w}{%
\pi\lambda_{\mathrm{eff}}}\frac{\pi s}{2w}\right) -G\left( \frac{\pi s}{2w}%
\right) }{\left( \frac{\pi\lambda_{\mathrm{eff}}}{w}\right) ^{2}-1}
\\
& +\frac{\pi}{2\left( \frac{\pi\lambda_{\mathrm{eff}}}{w}+1\right) }%
\sin\left( \frac{\pi s}{2w}\right) ,
\end{aligned}
\label{app:pair_pos_29}
\end{equation}
where%
\begin{equation}
G\left( u\right) =\cos u\,\mathrm{Ci}\left( u\right) +\sin u\left[ \mathrm{Si%
}\left( u\right) -\frac{\pi}{2}\right] .  \label{app:pair_pos_28}
\end{equation}

\subsection{Optimal pair position in the logarithmic interaction regime}

For the logarithmic interaction regime, 
\begin{equation}
\xi\ll2x\ll\lambda_{\mathrm{eff}},  \label{app:pair_pos_35}
\end{equation}
the vortex--antivortex interaction is 
\begin{equation}
U_{\mathrm{v\bar{v}}}\left( R\right) =-\frac{\phi_{0}^{2}}{8\pi^{2}\lambda_{%
\mathrm{eff}}}\ln\left( \frac{\lambda_{\mathrm{eff}}}{R}\right) .
\label{app:pair_pos_36}
\end{equation}
Hence 
\begin{equation}
\frac{dU_{\mathrm{v\bar{v}}}\left( R\right) }{dR}=\frac{\phi_{0}^{2}}{%
8\pi^{2}\lambda_{\mathrm{eff}}}\frac{1}{R}.  \label{app:pair_pos_37}
\end{equation}
At $R=2x$, 
\begin{equation}
\left. \frac{dU_{\mathrm{v\bar{v}}}\left( R\right) }{dR}\right\vert _{R=2x}=%
\frac{\phi_{0}^{2}}{16\pi^{2}\lambda_{\mathrm{eff}}x}.
\label{app:pair_pos_38}
\end{equation}
\cref{app:pair_pos_23} then becomes 
\begin{equation}
\frac{{\alpha_0}}{4\pi^{3}}\left[ J^{\prime}\left( w+x\right) +J^{\prime }\left(
w-x\right) \right] =\frac{1}{16\pi^{2}\lambda_{\mathrm{eff}}x}.
\label{app:pair_pos_40}
\end{equation}
Equivalently, 
\begin{equation}
\frac{J^{\prime}\left( w+x\right) +J^{\prime}\left( w-x\right) }{\frac {\pi}{%
2w}}=\frac{w}{2{\alpha_0}\lambda_{\mathrm{eff}}x}.  \label{app:pair_pos_41}
\end{equation}
This equation shows explicitly how the vortex--antivortex attraction shifts
the optimal position. In the absence of the vortex--antivortex 
interaction, the right-hand side vanishes. 
The attraction between the vortex and
antivortex gives a positive right-hand side and therefore moves the optimum
away from the extremum of the isolated vortex--wall potential. Thus, the
correct optimal position is determined by the balance between the force
exerted by the magnetoelectric wall potential and the attractive
vortex--antivortex force.

To make the equation dimensionless, we introduce 
\begin{equation}
u_{\ast}=\frac{x_{\ast}}{w},\qquad\Lambda=\frac{\lambda_{\mathrm{eff}}}{w}%
,\qquad A={\alpha_0}w.  \label{app:pair_pos_log_18}
\end{equation}
where $x_{\ast}$\ is the optimal position of a vortex. Then Eq.~(\ref%
{app:pair_pos_41}) becomes 
\begin{equation}
\frac{1}{2Au_{\ast}}=\frac{\lambda_{\mathrm{eff}}}{w}\frac{\left[ J^{\prime
}\left( w\left( 1+u_{\ast}\right) \right) +J^{\prime}\left( w\left(
1-u_{\ast}\right) \right) \right] }{\frac{\pi}{2w}}.
\label{app:pair_pos_log_20}
\end{equation}
Thus the numerical value of $x_{\ast}/w$ depends on both $\Lambda$ and $A$.

\begin{figure}[H]
\centering
\includegraphics[width=\columnwidth]{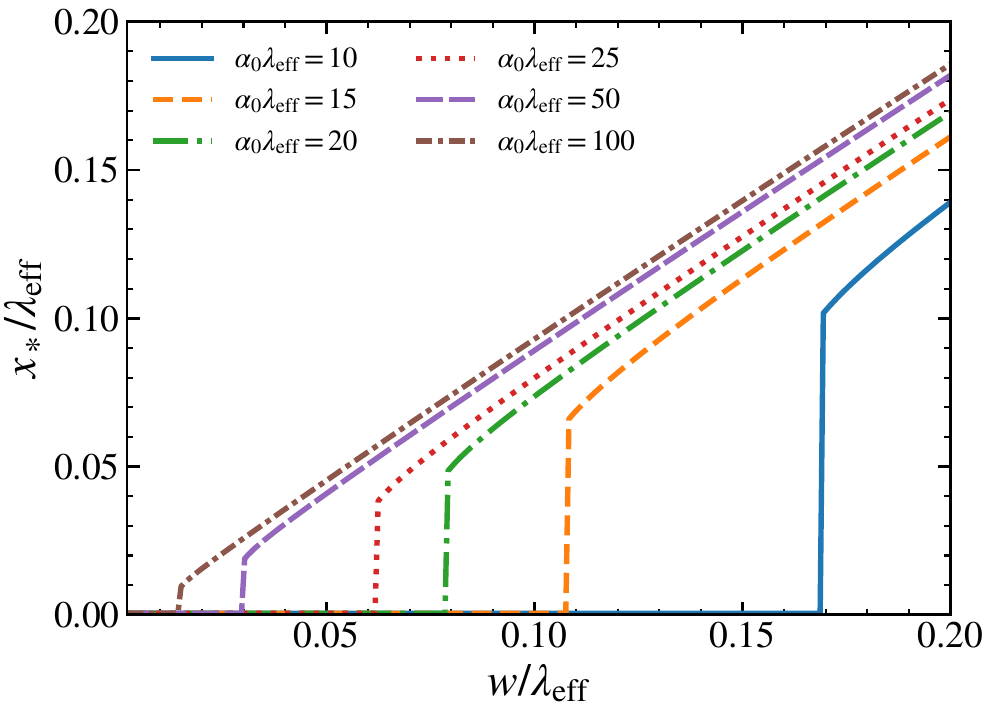}
\caption{ Dependence of the optimal vortex--antivortex position $x_{\ast}/\lambda_{\mathrm{eff}}$
on the ratio $\protect w/\lambda_{\mathrm{eff}}$ for several values of the
dimensionless magnetoelectric coupling $\protect\alpha_{0} \lambda_{\mathrm{eff}}$. }
\label{fig:0A}
\end{figure}

\begin{figure}[htbp]
\centering
\includegraphics[width=\columnwidth]{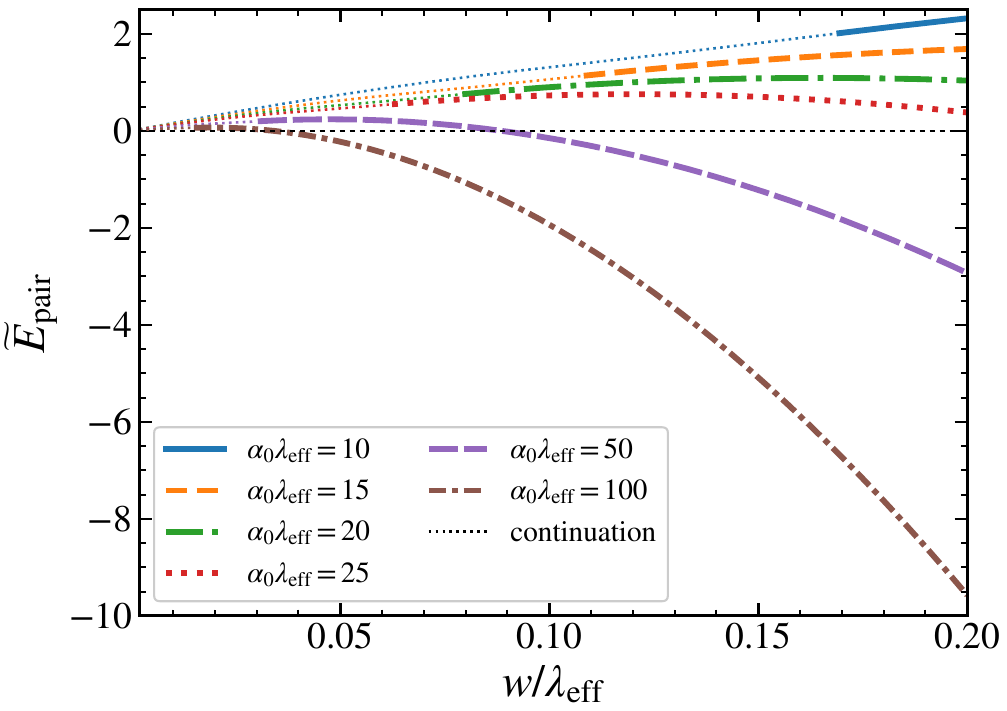}
\caption{Dependence of the dimensionless vortex--antivortex pair energy $%
\widetilde{E}_{\mathrm{pair}}(x_{\ast})$ on the ratio $\protect w/\lambda_{%
\mathrm{eff}}$ for several values of the magnetoelectric coupling strength 
$\protect\alpha_{0} \lambda_{%
\mathrm{eff}}$. The energy is evaluated at the equilibrium vortex
position $x_{\ast}$ determined from minimization of the full pair energy }
\label{fig:0B}
\end{figure}

\cref{fig:0A} shows the dependence of the optimal vortex--antivortex
position $x_{\ast }/\lambda _{\mathrm{eff}}$ on the ratio $w/\lambda _{%
\mathrm{eff}}$ for several values of the dimensionless magnetoelectric
coupling $\alpha_0 \lambda _{\mathrm{eff}}$. The position is determined from
the minimum of the full pair energy in the logarithmic interaction regime $%
\xi \ll 2x_{\ast }\ll \lambda _{\mathrm{eff}}$. Increasing $w/\lambda _{%
\mathrm{eff}}$ enhances the spatial range of the magnetoelectric interaction
and shifts the equilibrium vortex and antivortex positions toward the edges
of the N\'{e}el wall. Larger values of $\alpha_0 \lambda _{\mathrm{eff}}$
further stabilize configurations with vortex and antivortex localized near
opposite sides of the wall.
\cref{fig:0B} shows the dependence of the dimensionless
vortex--antivortex pair energy $\widetilde{E}_{\mathrm{pair}}\left( x_{\ast
}\right) $ on the ratio $w/\lambda _{\mathrm{eff}}$ for several values of
the magnetoelectric coupling strength $\alpha_{0} \lambda _{\mathrm{eff}}$. The
energy is evaluated at the equilibrium vortex position $x_{\ast }$
determined from minimization of the full pair energy.

\clearpage

\bibliography{references}